\documentclass[aps,pra,reprint,showpacs,showkeys,superscriptaddress,nofootinbib]{revtex4-1}

\usepackage{amsfonts,amssymb,mathrsfs,dsfont}
\usepackage{amsopn}
\usepackage{amsmath,amsthm}
\usepackage{enumitem}
\usepackage{float}
\usepackage{graphicx}
\usepackage{tikz} 
\usetikzlibrary{arrows}
\usepackage{hyperref}
\hypersetup{
   colorlinks,
   menucolor=black,
   linkcolor=black,
   citecolor=black,
   urlcolor=blue
}

\definecolor{darkgreen}{rgb}{0,0.5,0}
\definecolor{darkred}{cmyk}{0,1,1,0.4}

\DeclareMathOperator{\curl}{\mathrm{curl}}

\DeclareMathAlphabet{\mathpzc}{OT1}{pzc}{m}{it}

\newcommand\1{{\ensuremath {\mathds 1} }}
\newcommand{\C}{\mathbb{C}}

\newcommand{\N}{\mathbb{N}}

\newcommand{\R}{\mathbb{R}}
\newcommand{\Z}{\mathbb{Z}}
\newcommand{\bA}{\mathbf{A}}
\newcommand{\ba}{\mathbf{a}}

\newcommand{\be}{\mathbf{e}}
\newcommand{\bj}{\mathbf{j}}
\newcommand{\bJ}{\mathbf{J}}

\newcommand{\bx}{\mathbf{x}}

\newcommand{\by}{\mathbf{y}}

\newcommand{\cE}{\mathcal{E}}

\newcommand{\cJ}{\mathcal{J}}

\newcommand{\cO}{\mathcal{O}}

\newcommand{\sU}{\mathrm{U}}

\newcommand{\sym}{\mathrm{sym}}
\newcommand{\asym}{\mathrm{asym}}

\newcommand{\CLGN}{C_{\rm LGN}}

\newcommand{\cETF}{\cE^{\mathrm{TF}}}

\newcommand{\ETF}{E^{\mathrm{TF}}}

\newcommand{\rhoTF}{\tilde\varrho^{\mathrm{TF}}}

\newcommand{\Vh}{V_\mathrm{h}}

\newcommand{\mat}[1]{
  \begin{bmatrix}
    #1
  \end{bmatrix}
}

\newcommand{\norm}[1]{\left\lVert #1 \right\rVert}

\theoremstyle{plain}

\theoremstyle{definition}

\theoremstyle{remark}

\newtheorem*{ack}{Acknowledgments}

\begin{document}

\title{Nonlinear Landau levels in the almost-bosonic anyon gas}

\author{Alireza Ataei}
\affiliation{Department of Mathematics, Uppsala University, Box 524, SE-751 20, Uppsala, Sweden}
\author{Ask Ellingsen}
\affiliation{Department of Mathematics, Uppsala University, Box 524, SE-751 20, Uppsala, Sweden}
\author{Filippa Getzner}
\affiliation{Department of Mathematics, Uppsala University, Box 524, SE-751 20, Uppsala, Sweden}
\author{Th\'eotime Girardot}
\affiliation{Gran Sasso Science Institute, GSSI, 67100, L'Aquila, Italy}
\author{Douglas Lundholm}
\affiliation{Department of Mathematics, Uppsala University, Box 524, SE-751 20, Uppsala, Sweden}
\author{Dinh-Thi Nguyen}
\affiliation{Department of Mathematics, Uppsala University, Box 524, SE-751 20, Uppsala, Sweden}
\affiliation{Faculty of Mathematics and Computer Science, University of Science, Ho Chi Minh City, Vietnam}
\affiliation{Vietnam National University, Ho Chi Minh City, Vietnam}

\begin{abstract}
	We consider the quantitative description of a many-particle gas of
	interacting abelian anyons in the plane, confined 
    in a trapping potential.
	If the anyons are modeled as bosons with a magnetic flux attachment, 
	and if the total magnetic flux is small 
	compared to the 
	number of particles, then an average-field description becomes appropriate
	for the low-energy collective state of the gas.
	Namely, by means of a Hartree--Jastrow ansatz, 
	we derive a two-parameter 
	Chern--Simons--Schr\"odinger 
    energy functional 
	which extends the well-known Gross--Pitaevskii 
	/ 
    nonlinear Schr\"odinger 
    density functional theory to the magnetic (anyonic)
	self-interaction.
	One parameter determines the total number of 
	self-generated magnetic flux units in the system, 
	and the other the effective strength of 
    spin-orbit self-interaction. 
    This latter 
    interaction can be either attractive/focusing or repulsive/defocusing,
	and depends both on the intrinsic spin-orbit interaction 
	and the relative length scale of the flux profile of the anyons.
    Densities and energies of ground and excited states are studied
	analytically and numerically for a wide range of the parameters
    and align well with 
    a sequence of exact nonlinear Landau levels 
    describing 
    Jackiw--Pi self-dual solitons.
    With increasing flux, counter-rotating vortices are formed, enhancing the stability of the gas against collapse.
	Apart from clarifying the relations between various different anyon models 
	that have appeared in the literature,
	our analysis 
	sheds new light on the 
    many-anyon spectral problem,
	and also exemplifies a novel supersymmetry-breaking phenomenon.
\end{abstract}


\maketitle


\setlength\arraycolsep{2pt}
\def\arraystretch{1.2}

\section{Introduction}
\label{sec:intro}

Planar physics offers exciting possibilities by opening up for 
intermediate quantum statistics and anyons \cite{LeiMyr-77,GolMenSha-80,GolMenSha-81,Wilczek-82a,Wilczek-82b,AroSchWil-84,Wu-84a,GolMenSha-85},
with potentially groundbreaking applications in topological quantum information and computation
\cite{Kitaev-03,Lloyd-02,FreKitLarWan-03}.
However, 
despite recent advances in the experimental realization of individual anyons \cite{Bartolomei-etal-20,Nakamura-etal-20,Google-23,ReaDas-24},
a major obstacle to 
large-scale study of anyon physics
is our yet limited theoretical understanding of the basic properties of the anyon gas.
Compare, for example, the essential role 
\cite{DalGioPitStr-99,BaoCai-12,PitStr-16} 
of the Gross--Pitaevskii (GP) theory of the dilute interacting Bose gas \cite{Gross-61,Pitaevskii-61} for the eventual precise experimental study of Bose--Einstein condensation, more than three decades later
\cite{CorWie-nobel,Ketterle-nobel}.
Analogously, in order to reliably study collective properties of any large numbers of anyons emerging in experimental settings, there is a need for robust effective models for anyon gases, including a rigorous understanding of their range of validity and stability, as well as an availability of concrete energy and density signatures to compare experiments to.
Here then the actual many-body quantum statistics of anyons plays a crucial role.

The key to unraveling the fundamental relationship between exchange and exclusion statistics for anyons is their enigmatic $N$-body spectral problem, which is mathematically challenging even in the simplest possible (ideal and abelian) case 
\cite{CanGir-90,CanJoh-94,Khare-05}. 
Namely, while $N=2$ is analytically tractable,
exhibiting a simple interpolation 
between the properties (energy, density, etc.) of bosons and fermions 
\cite{LeiMyr-77,Wilczek-82b,Wu-84a},
only a small fraction of the spectrum is known analytically for $N \ge 3$ 
\cite{Wu-84b,Chou-91a,Chou-91b}.
The remaining part has been investigated using numerical methods for small systems
\cite{HatKohWu-91a,SpoVerZah-91,MurLawBraBha-91,SpoVerZah-92}.
In the early 1990s, the qualitative behavior of the ideal $N$-anyon spectrum in a harmonic trap was studied in the limit of large $N$ \cite{ChiSen-92,LiBhaMur-92}. 
It showed a highly nontrivial interpolation in terms of angular momentum, and included some quantitative approximations  by means of density functionals
in regimes close to the limits of bosons and fermions.
Around the same time, 
it was also realized that pointlike but singular (point-interacting) anyons can exist \cite{Girvin-etal-90,Grundberg-etal-91,ManTar-91,SenChi-92,BorSor-92,KimLee-97}, and may manifest a
``noninterpolating'' spectrum \cite{Chou-91a,MurLawBhaDat-92}, in contrast to the ideal or regular ones.
A possible physical interpretation
was given in terms of spin-orbit coupling and Pauli supersymmetry 
\cite{Grundberg-etal-91,ChoLeeLee-92,ComMasOuv-95,Mashkevich-96}.
Further, anyon superconductivity was proposed for certain fractions of the statistics parameter relative to fermions \cite{Laughlin-88,FetHanLau-89,CheWilWitHal-89}, and a Chern--Simons theory of statistics transmutation developed in the context of the fractional quantum Hall effect (FQHE) \cite{GirMac-87,ZhaHanKiv-89,LopFra-91,EzaHotIwa-92,TouHerHan-20}, 
which in a certain self-dual limit admits exact vortex soliton solutions
\cite{JacWei-90,JacPi-90b,JacPi-90a,Hagen-91,JacPi-91b,Jackiw-92,Dunne-92}.
Despite this intense development of the theory of anyons in the 1990s and extending into the 2000s
(see, e.g., the many books and substantial reviews \cite{Froehlich-90,Jackiw-90,Wilczek-90,Forte-92,IenLec-92,Lerda-92,Dunne-95,Dunne-99,Myrheim-99,DatMurVat-03,Khare-05,Ouvry-07,Nayak-etal-08,Stern-08,HorZha-09,HanHerSimVie-17}),
the central question concerning the precise relationship between exchange and exclusion for 2D anyons via their many-body spectrum remained unanswered.
For a recent encyclopedic overview of known properties of the 2D anyon gas, see \cite{Lundholm-23}.

In recent times, rigorous mathematical methods have been developed to investigate the stationary $N$-body ground state \cite{LunSol-13b,LunSol-13a,LunSol-14,LarLun-16,Lundholm-16,LunSei-17,LunQva-20,GirRou-23} and to derive effective and robust energy and density functional theories of Chern--Simons--Schr\"odinger (CSS) and Thomas--Fermi (TF) type 
\cite{LunRou-15,CorLunRou-17,CorLunRou-18,CorDubLunRou-19,GirRou-21,Girardot-21,GirLee-24,Visconti-25}.
In this work, we highlight and summarize some recent mathematical findings of physical importance for the abelian and almost-bosonic, interacting $N$-anyon problem \cite{AtaLunNgu-24,AtaGirLun-25}. They include the derivation of an effective average-field CSS model with precise relations for the physical parameters, its stability and supersymmetric properties, as well as the analysis of its emergent spectrum, which is numerically and analytically studied with the help of Jackiw--Pi vortex solitons or ``nonlinear Landau levels'' (NLLs). 
Our study embraces the universal many-body problem of interacting abelian anyons at the effective continuum level, starting from a sufficiently general Hamiltonian, independent from a particular experimental realization.
In particular, we investigate how the magnetic gauge interaction, which implements the statistics of anyons, gives rise, in the almost-bosonic regime, to an effective collective theory and how this theory interpolates between Gross--Pitaevskii-type behavior, self-generated magnetic states, and the NLLs associated with self-dual Chern--Simons vortices. This provides a tractable framework for studying many-anyon spectral properties that are inaccessible directly from the microscopic $N$-body Hamiltonian.

We approach the many-anyon spectral problem from the bosonic side, while also incorporating local interactions and scales akin to that of the dilute Bose gas
(for some recent progress from the fermionic side of the problem, see \cite{GirRou-21,Hu-etal-21,LevLunRou-25}).
Although the models for the anyon gas at different scales that we consider here could be considered vastly simplified compared to what may be the most realistic in current emergent settings, including present-day setups in quantum Hall or cold-atom experiments, we believe that they are sufficiently general and rich to capture at least two of the most essential features of interest: 
both the interpolating exclusion statistics of anyons, which acts on the global, macroscopic scale of a system, 
as well as some microscopic features acting on local, interparticle scales if the gas is dilute.
The intermediate model, describing a complex Schr\"odinger field self-coupled both locally and nonlocally via a Chern--Simons gauge field, with only two coupling parameters, is in a sense universal for planar systems \cite{Jackiw-90}. 
Using explicit relationships between the parameters, we can demonstrate a nontrivial interaction between these micro-, meso- and macroscopic scales.
In particular, this study represents a first step where we can exhibit an explict interpolation between ideal/regular/repulsive and singular/attractive models of anyons, and thereby quantitatively study the interplay between the uncertainty and exclusion principles for stability and for effective energy and density profiles of dilute anyon gases.
Further, the microscopic model we consider is not only of theoretical interest, since it has been shown that certain systems of at least two species of identical particles and broken time-reversal or orientation symmetry can yield emergent $N$-anyon Hamiltonians of a similar type, in quantum Hall setups
\cite{LunRou-16,Grass-etal-20,LamLunRou-22}
and in impurity setups
\cite{Yakaboylu-etal-19,ValWesOhb-20,Yakaboylu-etal-20,Brooks-etal-21}.

We finish this introduction with an outline of the present work.
Our approach is to start from a microscopic description of an emergent $N$-body anyon gas, with a couple of parameters to control its essential physical properties, such as the microscopic extent of the anyons and the strength of their local interactions. The explicit Hamiltonian is introduced and explained in Section~\ref{sec:micro}. 
In Section~\ref{sec:meso} we then introduce the relevant effective model on mesoscopic scales relative to the typical interparticle spacing in the gas, as well as outline its derivation from the microscopic Hamiltonian, assuming that the gas is sufficiently dilute and that the total amount of (self-generated) magnetic flux of the anyons is relatively small, i.e.\ in a regime sufficiently close to bosons.
Finally, in Section~\ref{sec:macro} we consider the macroscopic properties of such a gas on the scale of a harmonic trap, and investigate using analytical and numerical tools how the underlying microscopic and derived mesoscopic parameters influence the macroscopic energy and density distributions quantitatively. 
Conclusions and outlooks are given in Section~\ref{sec:outlooks}.
In the appendices we have gathered analytical and numerical supplemental material that has been summarized in the main text and in the figures.

\section{Microscopic anyon Hamiltonian}
\label{sec:micro}

For a microscopic description of the anyon gas,
we consider the following $N$-anyon Hamiltonian 
(see \cite{Mashkevich-96,AtaGirLun-25,Lundholm-23} and below):
\begin{equation}\label{eq:H}
    H_N 
    := \sum_{j=1}^{N} \left[ 
    	\left(-i\nabla_{\bx_j}+ \alpha \bA^R_j\right)^{2}
	+ \frac{g}{2} \alpha B^R_j + V(\bx_j)
	\right].
\end{equation}
Here, $\bx_j \in \R^2$ denotes the position in the plane of the $j$:th particle (in an arbitrary numbering since the particles are identical),
and $\alpha$ is the number of magnetic flux units attached to each particle, 
i.e.\ the statistics parameter of the anyons. 
The planar vector fields
\begin{equation}\label{eq:AR}
    \bA^R_j := \sum_{k\neq j}^N \frac{(\bx_j -\bx_k)^{\perp}}{|\bx_j -\bx_k|_R^2}, 
\end{equation}
defined with $|\bx|_R:=\max\{R,|\bx|\}$ and $\bx^{\perp}:=(-x_2,x_1)$, 
represent magnetic vector
potentials in Coulomb gauge with magnetic field  
\begin{equation}\label{eq:BR}
    B^R_j := \curl \bA^R_j = \frac{2}{R^2} \sum_{k \neq j} \1_{D(\bx_k,R)}(\bx_j).
\end{equation}
This magnetic field is supported uniformly on disks of radius $R$ each with a single unit $2\pi$ of flux centered at $\bx_k$, as seen by particle $j$.
This flux profile is not the most general or realistic, but a sufficently good approximation to emergent flux attachment in some models (see below).
We also allow pointlike anyons by taking $R \to 0$ 
and $\1_{D(\bx_k,R)}/(\pi R^2) \to \delta_{\bx_k}$, 
the Dirac measure.
The coupling parameter $g \in \R$ to the magnetic field will be detailed below.
Lastly, $V$ is a regular one-body trapping potential, 
i.e.\ $V(\bx) \to \infty$ as $\bx \to \infty$,
which acts to confine the gas to a suitable domain.
For definiteness and simplicity,
here we will consider the
harmonic trap $\Vh(\bx) := \frac{1}{2}m\omega^2|\bx|^2$, 
although this can be easily generalized.
We normalize units 
$\hbar = 2m = \frac{\omega}{2} = 1$.
See Appendix~\ref{app:notation} for a more detailed recap of some standard notation.

In our conventions used here,
the Hamiltonian \eqref{eq:H} 
acts on symmetric wave functions 
$\Psi_N \in L^2_\sym(\R^{2N})$, so that 
$\alpha=0$ corresponds to ideal bosons and 
$\alpha=1$, $R=0$ to ideal, \emph{spinless} fermions.
Thus, in the magnetic-gauge representation used here, the exchange statistics are encoded in the statistical gauge fields $\bA^R_j$ rather than in the permutation symmetry of the many-body wave function.
Indeed, in the pointlike limit $R=0$, we write 
the wave function
in the anyon-gauge representation
\begin{equation}\label{eq:stat-transm}
    \Phi_N(\bx_1,\ldots,\bx_N) 
    := \prod_{j<k}e^{i\alpha\theta_{jk}} \Psi_N(\bx_1,\ldots,\bx_N),
\end{equation}
where
\begin{equation}\label{eq:stat-transm-arg}
    \theta_{jk}
    := \arg\frac{z_{j}-z_{k}}{|z_{j}-z_{k}|}
\end{equation}
denotes the relative angle in the plane between particles $j$ and $k$, 
represented in the complex plane by positions $z_j,z_k \in \C$.
We may then formally exhibit the statistics transmutation
\begin{equation}\label{eq:stat-transm-H}
    \langle \Psi_N | H_N | \Psi_N \rangle
    = \langle \Phi_N | \tilde{H}_N | \Phi_N \rangle,
\end{equation}
with the (formal; see below) Hamiltonian
\begin{equation}\label{eq:H-transm}
    \tilde{H}_N
    := \sum_{j=1}^{N} \left[ 
        -\nabla_{\bx_j}^2
	+ \frac{g}{2} 2\pi\alpha \sum_{k \neq j} \delta(\bx_j-\bx_k) + V(\bx_j)
	\right].
\end{equation}
In general $\Phi_N$ is multivalued, and an elementary counterclockwise simple exchange $j \leftrightarrow k$ yields the abelian anyonic phase $e^{i\alpha\pi}$.
In the case that $\alpha=1$, then $\Phi_N \in L^2_\asym(\R^{2N})$ describes an antisymmetric state, as appropriate for $N$ fermions projected into a single spin component.

\subsection{Spin-orbit interaction and supersymmetry}

In fact, we generally interpret $g$ in \eqref{eq:H} (respectively \eqref{eq:H-transm}) as a spin-orbit coupling parameter.
Namely, consider a one-body Pauli operator in the two-dimensional plane with external magnetic potential $\bA = (A_1,A_2)$ and field $B = \curl \bA$, acting on a spin-1/2 system $L^2(\R^2) \otimes \C^2$:
\begin{equation}\label{eq:Pauli}
    H^{\rm Pauli}_{\bA} = \bigl(-i\nabla + \bA(\bx)\bigr)^2 \otimes \1_{\C^2} + g B(\bx) \mat{+1/2 & 0 \\ 0 & -1/2}
\end{equation}
(in our units, the Bohr magneton is $\mu_{\rm B} = e\hbar/(2m) = 1$).
For exactly the values $g=\pm 2$, this expression exhibits ``supersymmetry'', 
by completing the square of a 2D Dirac operator:
\begin{equation}\label{eq:Pauli-susy}
    H^{\rm Pauli}_{\bA} = \left[ \sum_{k=1,2} \sigma_k (-i\partial_k + A_k) \right]^2 \ge 0,
\end{equation}
where $\sigma_k \in \C^{2 \times 2}$ are the Pauli matrices such that 
$\sigma_1\sigma_2\sigma_3 = \pm i\1$.
In this interpretation, the Dirac operator is the supercharge operator, i.e.\ the generator of supersymmetry, and the space of supersymmetric states (those annihilated by the supercharge) is exactly the space of zero-energy states of the Hamiltonian
(see, for example, \cite{deCRit-83,Jackiw-84,Lundholm-PhD} for more background, detail, and examples of such supersymmetric quantum mechanics).

Thus, 
considering $B=\alpha B^R_j$ for each particle $j$,
the interacting $N$-anyon Hamiltonian
\eqref{eq:H}
can indeed be interpreted as the projection to one of the spin components of such a spin-orbit coupled Pauli operator in the field of the other particles. 
In general, $g \in \R$ is interpreted as the magnetic moment or the gyromagnetic ratio (the ``Land\'e $g$-factor''), and was discussed to some extent in \cite{Grundberg-etal-91,ComMasOuv-95,Mashkevich-96} (see also \cite{JacPi-90a,ChoLeeLee-92}). However, we are not aware of any discussion in the literature which fixes its precise value in an emergent anyon model, and shall therefore leave it as one of the main microscopic parameters of our model and study its effect on the emergent mesoscopic model, as well as in the emergent macroscopic model. 

\subsection{Point-interacting anyons}

Note that, even in the pointlike limit $R=0$ at fixed $N$ and $\alpha$, 
there are effectively a range of different anyon types
specified by $g$,
where $g=\pm 2$ defines {\bf ``ideal anyons''} (regular/hard)
and {\bf ``kreinyons''} (singular/soft), 
respectively. 

Namely, due to the singular nature of the anyonic gauge vector potential $\bA_j^{R=0}$, the two-body Hamiltonian of pointlike anyons for $0<\alpha<1$ admits a one-parameter family of different self-adjoint extensions, or point interactions, corresponding to the boundary conditions
\begin{equation}\label{eq:point-bc}
    \Psi_{N=2}(r) \sim a r^\alpha + b r^{-\alpha},
    \qquad \text{as} \ r = |\bx_1-\bx_2| \to 0,
\end{equation}
for $a,b \in \C$ not both zero. 
Most of these extensions have a scale and some of those admit negative bound states, but 
exactly two of the extensions are both non-negative and scale-covariant. 
These are in the mathematics literature known as the Friedrichs extension ($b=0$, with the largest energy or the most positive) 
and the Krein extension ($a=0$, with the smallest energy or the least positive);
cf. \cite{Simon-79b,AloSim-80}. 
In the physics literature, these anyons are usually 
(despite some ambiguity)
referred to as hard-core/repulsive/regular, respectively soft-core/attractive/singular; see \cite{Girvin-etal-90,Grundberg-etal-91,ManTar-91,SenChi-92,BorSor-92,Mashkevich-96}.

Further, in the well-studied pointlike $N$-anyon problem \eqref{eq:H-transm} with harmonic confinement, $V = \Vh$, both of these types of anyons can be found among the exact (generalized) eigenfunctions and are then also referred to as the interpolating solutions (corresponding to ideal anyons; see \cite{Chou-91b,Chou-91a,ChiSen-92}), respectively the noninterpolating solutions (see \cite{Chou-91a,MurLawBhaDat-92}). 
Their energies at fixed $N$ vary linearly in $\alpha$.
For example, for small enough $\alpha N$, the exact $N$-anyon ground state is analytically known:
\begin{equation}\label{eq:true-gs}
    \Psi_N^{\pm} \propto \prod_{j<k} |\bx_j-\bx_k|^{\pm\alpha} e^{-\frac{1}{2}\sum_j|\bx_j|^2},
\end{equation}
with energy
\begin{equation}\label{eq:true-gs-energy}
    \langle \Psi_N^{\pm} | H_N | \Psi_N^{\pm} \rangle 
    = 2N \pm \alpha N(N-1).
\end{equation}
The regular family $\Psi_N^+$ corresponds to ideal anyons, while the singular family $\Psi_N^-$ we thus term ``kreinyons''
(for ``scale-covariant non-negative singular pointlike anyon extensions of the Krein type''),
attempting to make the terminology less ambiguous.
Also, by studying the local spin-orbit interaction in the limit $R \to 0$ for $N=2$, it is clear that $g=2$ corresponds to ideal anyons and $g=-2$ to such ``kreinyons'' \cite{Mashkevich-96,AtaGirLun-25}.
In the case $\alpha=0$ of two bosons in the plane, the Friedrichs and Krein extensions coincide (free, ideal bosons) and any other extension has a negative bound state (attractive point interaction).
In the case $\alpha=1$ of fermions, the possibility for point interaction drops out, since the singular state is no longer in the Hilbert space.
Thus, $g$ is then irrelevant and pointlike spinless fermions are ideal (and regular/hard).
See a longer discussion in \cite{Lundholm-23,AtaGirLun-25} for more details.

The rigorous mathematical theory of $N$-body point interacting Hamiltonians of the type \eqref{eq:H-transm} is still in its infancy (see, e.g., \cite{DelFigTet-94,CorOdd-18,BorCorFer-24}),
and we do not claim to give a complete classification of anyons here.
However, by first taking $R>0$, and then $R \to 0$ at fixed $g$, we can avoid any formal difficulties with defining expressions such as \eqref{eq:H-transm}.
Also note that actual emergent anyons are indeed expected to have a finite size $R>0$, such as on the order of the magnetic length of a strong external field; see, e.g., \cite{LunRou-16,Yakaboylu-etal-19,LamLunRou-22}. Since that is typically much smaller than the domain of an experiment, it is legitimate to consider a pointlike limit $R \to 0$ if the relevant microscopic information can be retained,
and this is precisely the role of the spin-orbit coupling parameter $g$ in this study. 

The conventional Berry phase approach to emergent anyons \cite{AroSchWil-84} is not sufficiently precise to determine the microscopic anyon model
\cite{Forte-91,LunRou-16,LamLunRou-22},
and in particular the parameter $g$, which takes an essential role in the Hamiltonian description.
However, for tracer particles coupled to Laughlin quasiholes in the FQHE, an explicit emergent anyonic Hamiltonian of the form \eqref{eq:H} has indeed been conjectured in \cite[Conjecture~1.2 and Section~5]{LamLunRou-22}, with the assumption that the local interactions between tracer particles are sufficiently repulsive, i.e.\ $g>0$ is sufficiently large.
The limit $g \to +\infty$ represents impenetrably hard disks.
In the FQHE setting it may be a challenge to reach the relatively small values of $\alpha$ considered here, and an alternative and more feasible setting could instead be polarons in rotating ultra-cold atomic gases \cite{Yakaboylu-etal-19}. 
Finally, we note that in the emergent description it is also important to study the inclusion of external magnetic fields, however we have chosen to postpone such a study (see \cite{Scribano-26} for some recent initial results) and to here focus on the purely self-generated statistical effects, for clarity.

\section{Mesoscopic CSS average-field functional}
\label{sec:meso}

\subsection{Chern--Simons--Schr\"odinger model}

\begin{figure}
	\centering
	\includegraphics[scale=0.35,clip,trim=1.3cm 0.5cm 1.3cm 0.7cm]{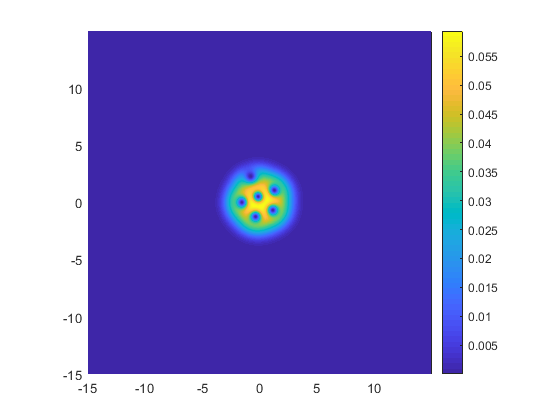}%
	\includegraphics[scale=0.35,clip,trim=1.3cm 0.5cm 1.3cm 0.7cm]{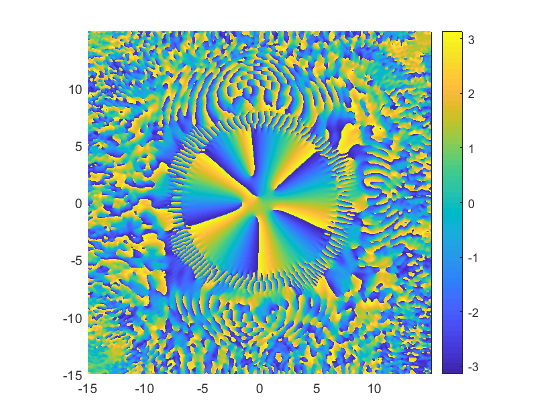}
	\caption{Density and phase of an approximate minimizer $\phi$
	for $E(\beta=10,\gamma=20\pi,\Vh) \approx 9.066$.}
	\label{fig:density-phase}
\end{figure}

Lacking a complete spectral resolution of the Schr\"odinger Hamiltonian \eqref{eq:H} for $0<\alpha<1$,
the microscopic description becomes analytically and numerically intractable for large particle numbers. 
 However, in the almost-bosonic scaling
\begin{equation}\label{eq:a-b-limit}
    N\gg1,
    \qquad
    \alpha N\to\beta=O(1),
\end{equation}
the total statistical flux, denoted $\beta$, 
is considered to be of order unity while 
the flux attached to each individual particle is small. 
In this limit the individual fluxes can be replaced by a smooth magnetic field proportional to the density of the gas.
This is known as the average-field approximation in the literature \cite{Wilczek-90}.
On intermediate length scales $\ell$,
\begin{equation}\label{eq:meso-scale}
    R \ll \ell \ll \omega^{-1/2},
\end{equation}
the interacting anyon gas is then described, at the level of its leading energy per particle, by the following Chern--Simons--Schr\"odinger (CSS) energy functional:
\begin{align}\label{eq:CSS-func}
	&\cE_{\beta,\gamma,V}[\phi] 
	:=   \\
    &\int_{\R^2} \left[
	\bigl| (-i\nabla + \beta\bA_\varrho) \phi(\bx) \bigr|^2
		+ \gamma |\phi(\bx)|^4
		+ V(\bx) |\phi(\bx)|^2 \right] \mathrm{d}\bx. \nonumber
\end{align}
Here
$\phi \in L^2(\R^2)$ is a one-body collective state,
the CSS wave function 
(normalized, regular and singlevalued; 
see Fig.~\ref{fig:density-phase} for an example),
\begin{align*}
  \bA_\varrho(\bx) :=  \int_{\R^2} \frac{(\bx-\by)^{\perp}}{|\bx-\by|^2} \varrho(\by) \, \mathrm{d} \by
\end{align*}
is a magnetic vector potential in Coulomb gauge with the self-consistently generated magnetic field 
\begin{equation}\label{eq:Brho}
    B_\varrho := \curl \bA_\varrho = 2\pi\varrho,
\end{equation}
with density distribution
$\varrho = |\phi|^2$
(see Appendix~\ref{app:notation} for details),
$V(\bx)$ the same trapping potential as in \eqref{eq:H},
and $\gamma$ is an effective coupling parameter, detailed below.

To better understand the effective functional \eqref{eq:CSS-func}, it is helpful to relate it to the standard Chern--Simons--Schr\"odinger formulation. At the level of a time-dependent non-relativistic field theory, one introduces a scalar self-interacting complex matter field $\phi(t,\mathbf x)$, 
trapped into 
the stationary electric potential $V(\bx)$,
and minimally coupled to a Chern--Simons gauge field $a_\mu=(a_0,\mathbf a)$; 
see \cite{Semenoff-88,JacPi-90b,JacPi-90a,Dunne-95}.
The Lagrangian density of the system is then $\mathcal{L}=\mathcal{L}_{\mathrm{NLS}}+\mathcal{L}_{\mathrm{CS}}$ 
(nonlinear Schr\"odinger plus Chern--Simons; with no Maxwell term)
such that
\begin{align*}
    \mathcal{L}_{\mathrm{NLS}}&:= \bar{\phi}(i\partial_t - a_0)\phi
    - \bar{\phi}(-i\nabla + \ba)^2\phi
    - V|\phi|^2 - \gamma|\phi|^4,\\
    \mathcal{L}_{\mathrm{CS}}&:=\frac{k}{2} \epsilon^{\mu\nu\lambda} a_\mu \partial_\nu a_\lambda .
\end{align*}
In the above, $\epsilon^{\mu\nu\lambda}$ is the antisymmetric tensor such that $\epsilon^{012}=1$,
and $k$ the CS coupling parameter.
We have included the quartic term $\gamma|\phi|^4$ 
for the usual, effectively \emph{local}, two-body nonlinear Schr\"odinger interaction (nonrelativistic $\phi^4$ theory) with scalar coupling $\gamma$.
Varying the action $\mathcal{S}[\phi,a_{\mu}]=\int \mathcal{L}[\phi,a_{\mu}]\mathrm{d}\bx\mathrm{d}t$ along the field $a_0$ and $a_j$ respectively provides modified Gauss and Amp\`ere laws for the magnetic and electric fields $(b,\be)$ related to the matter field via 
\begin{equation}\label{eq:be}
b= k^{-1}
|\phi|^2,\quad \text{and}\quad \be =k^{-1} \bj^{\perp}[\phi],
\end{equation}
where $\bj[\phi]$ is the matter current.
Crucially, this brings about additional nonlinear and \emph{nonlocal} couplings to the density $\varrho = |\phi|^2$.
Thus, having solved the constraints and
by setting $\ba = \beta\bA_\varrho$, which corresponds to working in the Coulomb gauge, we can apply the Legendre transform described in \cite{Semenoff-88} and recover our effective Hamiltonian \eqref{eq:CSS-func} 
with $k=(2\pi\beta)^{-1}$ and $\phi$ the remaining variable. 
In the combined theory, the special values
$\gamma = \pm2\pi\beta$ 
simplify the equations of motion
and correspond to self-dual solutions
\cite{JacPi-90a} 
(again related to the supersymmetry of the Pauli Hamiltonian; see below;
in \cite[Chapter~II]{Dunne-95}, \eqref{eq:CSS-func} is simply referred to as the Pauli energy).

We denote the ground-state energy of \eqref{eq:CSS-func}, i.e. its infimum over regular $\phi$ subject to the normalization constraint
$\norm{\phi}=1$, by $E(\beta,\gamma,V)$.
For $\beta=0$, \eqref{eq:CSS-func} reduces to the 
Gross--Pitaevskii (GP) energy functional for interacting bosons in zero magnetic field,
which indeed is well known to be an accurate description for the Bose gas in the dilute limit \cite{DalGioPitStr-99,LieSeiSolYng-05}.

The Euler--Lagrange equation for a minimizing ground state or a stationary point 
$\phi$ of \eqref{eq:CSS-func} is the CSS equation
\begin{multline}\label{eq:CSS-EL}
    \left[
    (-i\nabla + \beta\bA_\varrho)^2 
    - 2\beta\nabla^\perp (\log |\cdot|) * \cJ_\phi
    + 2\gamma\varrho + V
    \right]\!\phi\,(\bx) \\
    = \mu \phi(\bx),
\end{multline}
where we defined the gauge equivariant current
$$
    \cJ_\phi 
    := \frac{1}{2} \overline{\phi} (-i\nabla + \beta\bA_\varrho) \phi + \text{h.c.}
$$
and $\mu = \mu(\phi) \in \R$ satisfies
\begin{equation}\label{eq:CSS-EL-lambda}
    \mu = 
    2\cE_{\beta,\gamma,V}[\phi]
    - \int_{\R^2} \left[|\nabla\phi|^2 + V\varrho - \beta^2 \left|\bA_\varrho\right|^2\varrho\right]\!(\bx)\,\mathrm{d}\bx;
\end{equation}
see \cite[Appendix]{CorLunRou-17} and \cite[Remark~3.11]{AtaLunNgu-24} for derivations.

\subsection{Derivation of the mesoscopic model}

The field-theoretic correspondence based on the Lagrangian $\mathcal{L}$ is 
useful for the interpretation of the effective model. However, it
should not be confused with the microscopic derivation of the effective functional \eqref{eq:CSS-func}. In our setting, the CSS functional is obtained by guessing a sufficiently precise ground state vector for the microscopic Hamiltonian \eqref{eq:H} in the almost-bosonic limit \eqref{eq:a-b-limit}. The guess leads to the Hartree--Jastrow class \eqref{eq:HJ-ansatz} made of a collective one-body state, while the Jastrow factor accounts for the short-range two-body correlations generated by the statistical interaction.

To approximate the low-energy stationary states of the anyon gas,
we suggest a Hartree--Jastrow ansatz for \eqref{eq:H}:
\begin{equation}\label{eq:HJ-ansatz}
	\Psi_N := C_N \prod_{i<j} f_{R/\ell}(|\bx_i-\bx_j|/\ell) \prod_{k=1}^N \phi(\bx_k),
\end{equation}
where $C_N > 0$ is a normalization constant s.t. $\norm{\Psi_N}=1$, 
$\phi \in L^2(\R^2)$ a smooth variational one-body state (CSS wave function),
and
$$
    f_s(r) := \begin{cases}
	\frac{4 s^{\alpha} }{2\left(1+s^{2\alpha}\right) + g\left(1-s^{2\alpha}\right)}, 
	\quad &\textup{if } 0 \le r \le s,\\
				\frac{(2+g) r^{\alpha} + (2-g) s^{2\alpha} r^{-\alpha} }{2\left(1+ s^{2\alpha}\right) + g\left(1-s^{2\alpha}\right)}, 
        \quad &\textup{if } s \le r \le 1,\\
        1, \quad &\textup{if } r \ge 1.
    \end{cases}
$$
Here $s=R/\ell$ is a relative length scale with $\ell$ on the order of the interparticle distance, and as in \eqref{eq:meso-scale}.

Intuitively, for $\alpha$ sufficiently small, we expect the bosons to form a Bose--Einstein condensate in a collective state $\phi$. However, due to the singular nature of the magnetic interactions as $R \to 0$, there needs to be a control of at least two-body correlations close to the diagonals of the configuration space; compare, e.g., the exact ground states \eqref{eq:true-gs}.
Our choice of $f_s$, resembling \eqref{eq:point-bc}, approximates the optimal such correlations.
The ansatz \eqref{eq:HJ-ansatz} thus describes a regime in which the gas is sufficiently dilute to separate the two- and the many-body effects on different scales. 
A Jastrow-type two-body correlation structure encodes the short-range effects up to interparticle distance or mesoscopic scale.
A Hartree-type condensate function averages any long-range many-particle effects, 
including those of a quantum-statistical nature.

Indeed, in \cite{AtaGirLun-25} it has been shown rigorously that, if the statistics parameter satisfies \eqref{eq:a-b-limit},
and the relative extent of anyons scales as
\begin{equation}\label{eq:rel-scale}
    0 \le R = \cO(e^{-\sigma N}) \ll \ell = \cO(N^{-\tau}),
\end{equation}
where $\beta>0$, $\sigma \ge 0$, $\tau > 2$, $g \ge 0$ are independent of $N$, then the energy per particle of \eqref{eq:HJ-ansatz} converges to the CSS energy:
\begin{equation}\label{eq:HJ-limit}
	\lim_{N \to \infty} N^{-1} \langle \Psi_N | H_N 
    | \Psi_N \rangle
	= \cE_{\beta,\gamma,V}[\phi],
\end{equation}
with the effective coupling parameter
\begin{equation}\label{eq:gamma-eff}
	\gamma = 2\pi \beta \frac{1+\frac{g}{2} - (1-\frac{g}{2}) e^{-2\beta\sigma}}{1+\frac{g}{2} + (1-\frac{g}{2}) e^{-2\beta\sigma}}.
\end{equation}
The two effective parameters $\beta$ and $\gamma$ have distinct microscopic origins. The parameter $\beta$ records the total statistical flux accumulated by the $N$ particles, $\beta\simeq\alpha N$, whereas $\gamma$ is generated by the short-distance two-body correlations encoded in the Jastrow factor $f_s$ (detailed below). 
Thus \eqref{eq:gamma-eff} provides the microscopic-to-mesoscopic parameter map: the statistical gauge interaction determines the self-generated magnetic coupling through $\beta$, while the short-range scattering problem determines the effective local nonlinear interaction through $\gamma$.

Note that, 
in the ``supersymmetric'' case $g=\pm 2$ this yields exactly $\gamma=\pm 2\pi\beta$ for all relative scales 
$\sigma \sim N^{-1}\log \frac{\ell}{R}$.
In the relatively dense limit $\sigma \to 0$ it becomes $\gamma=g\pi\beta$, whereas 
in the extremely dilute limit $\sigma \to +\infty$, $g > -2$ it is $\gamma=2\pi\beta$. 
We can also notice that in the non-interacting/``spinless'' limit $g \to 0$ it is $\gamma=2\pi\beta\tanh(\beta\sigma)$, 
and that $\gamma=2\pi\beta\coth(\beta\sigma)$ in the case of hard disks $g \to +\infty$, $\sigma>0$.
This extends the previously studied mean-field/Hartree case $\sigma=0=g$ (hence $\gamma=0$), $f_s=1$, for which matching lower bounds for the ground-state energies are also available
\cite{LunRou-15,Girardot-21,Visconti-25}.

\begin{figure}
	\centering
	\begin{tikzpicture}[domain=-2:12,scale=0.55]
	    \draw[very thin,color=gray] (-2,-7) grid (12,2);
	    \draw[->] (-2,0) -- (12.5,0) node[right] {$\beta$};
	    \draw[->] (0,-7.2) -- (0,2.3) node[left] {$\gamma$};
	    \draw[thin,dashed,color=blue,domain=-2:12.5,samples=50] plot (\x,{-0.93});
	    \draw[thin,dashed,color=blue,domain=-2:4.5,samples=50] plot (\x,{0.5*\x});
	    \draw[thin,dashed,color=blue,domain=-2:4,samples=50] plot (\x,{-0.5*\x});
	    \draw[thin,dotted,color=black,domain=-2:12.5,samples=50] plot (\x,{-0.5*9/10*\x}); 
	    \draw[very thick,dashed,color=red,domain=-2:4,samples=50] plot (\x,{-0.5*abs(\x) - 0.93*(4-abs(\x))^2/16});
	    \draw[very thick,color=red,domain=4:12.5,samples=50] plot (\x,-\x/2);
		\draw [fill,color=red] (0,-0.93) circle [radius=0.09];
		\draw [fill,color=darkgreen] (4,-2) circle [radius=0.09];
		\draw [fill,color=darkgreen] (8,-4) circle [radius=0.09];
		\draw [fill,color=darkgreen] (12,-6) circle [radius=0.09];
	    \node[above left] at (0,0) {$0$};
	    \node[above] at (2,0) {$1$};
	    \node[above] at (4,0) {$2$};
	    \node[above] at (6,0) {$3$};
	    \node[above] at (8,0) {$4$};
	    \node[above] at (10,0) {$5$};
	    \node[above] at (12,0) {$6$};
	    \node[left] at (0,-2) {$-4\pi$};
	    \node[left] at (0,-4) {$-8\pi$};
	    \node[left] at (0,-6) {$-12\pi$};
	    \node[below right] at (11.5,-0.9) {$-\CLGN\hspace{-1cm}$};
	    \node[below right] at (12,-6.2) {$-2\pi\beta\hspace{-1cm}$};
	    \node[above right] at (4.4,1.9) {$+2\pi\beta$};
	    \node[above right] at (11,-5.1) {$\kappa=-0.9$};
		\node [below right] at (9,-2) {stability};
		\node [below right] at (2,-5) {instability};
	\end{tikzpicture}
	\caption{Stability holds if and only if $E(\beta,\gamma,V) > -\infty$, which is the region above the red curve.
	NLLs exist at $\beta \in 2\Z \setminus \{0\}$ (green dots) with the critical self-dual coupling 
	$\gamma = -2\pi|\beta|$.}
	\label{fig:stability}
\end{figure}

On top of \eqref{eq:HJ-limit},
we also have a convergence (in the $L^1$ sense) of the one-particle density of $\Psi_N$ to $\varrho=|\phi|^2$:
\begin{equation}\label{eq:HJ-limit-dens}
	\lim_{N \to \infty} \int_{\R^{2(N-1)}} \left|\Psi_N(\bx,\bx_2,\ldots,\bx_N)\right|^2 \mathrm{d}\bx_2 \ldots \mathrm{d}\bx_N
	= \varrho(\bx).
\end{equation}
Furthermore, in the cases that matching lower bounds have been proven, there is generally a convergence (in an appropriate sense for one-body density matrices) of approximate ground states of \eqref{eq:H} into the corresponding manifold of CSS ground states of \eqref{eq:CSS-func} \cite{LunRou-15,Girardot-21,Visconti-25}.
We conjecture that the convergence of ground states holds in much greater generality extending to $\gamma \neq 0$ in \eqref{eq:gamma-eff}.

The derivation of the precise 
function $f_s$ and of
formula \eqref{eq:gamma-eff} was motivated by a previous result by Mashkevich \cite{Mashkevich-96} concerning the asymptotic two-body energy.
Namely,
we define a relative s-wave ``scattering energy'' $E_2$, coming from the minimization of $H_{N=2}$ using the center of mass coordinates, as the infimum of
$$
    \int_{D(0,1)} \left[ |\nabla f|^2 
    + \alpha^2 \frac{|\bx|^2}{\max\{R,|\bx|\}^4} |f|^2 
    + \alpha g \frac{\1_{D(0,R)} }{R^{2}}|f|^2 \right],
$$
with $f \in L^2$ regular on the unit disk with $f|_{r=1}=1$.
Then our function $f_s$
captures exactly this scattering energy $E_2$
to order $\alpha^2$ in the limit, 
and its contribution to the total $N$-body energy per particle,
with $\binom{N}{2}$ pairs and relative mass 2,
is to leading order as $N \to \infty$
$$(N-1)E_2 \simeq \gamma$$ 
as in \eqref{eq:gamma-eff}; see \cite{AtaGirLun-25}.
The long-range interaction makes the scattering length and energy structurally different from the one for dilute Bose gases \cite{LieSeiSolYng-05,Solovej-25} usually defined for compactly supported interactions. 
We can, however, notice that the  long ranged anyonic vector interaction is suppressed by small $\alpha$, 
while the spin-orbit scalar interaction is relatively short ranged.

\subsection{Stability and supersymmetry}

For each $\beta$ there is a critical attractive coupling $\gamma$ for the CSS model \eqref{eq:CSS-func} beyond which the gas collapses.
We denote this critical absolute value by $\gamma_*(\beta)$.
Thus, for any $(\beta,\gamma)$ s.t. $\gamma \ge -\gamma_*(\beta)$, we have $E(\beta,\gamma,V) > -\infty$ (stability), while if  $\gamma < -\gamma_*(\beta)$ then $E(\beta,\gamma,V) = -\infty$ (instability); see Fig.~\ref{fig:stability}.
By scaling, it is sufficient to consider $V=0$ and $E(\beta,\gamma,0) = 0$ resp. $-\infty$, and the critical coupling $\gamma_*(\beta)$ is exactly the smallest (by infimum) ratio between the magnetic-kinetic self-energy $\cE_{\beta,0,0}[\phi]$ and the scalar/electrostatic self-energy $\int |\phi|^4$ among all normalized and regular $\phi$. 
Note the complex conjugation symmetry $\phi \to \overline{\phi}$, $\beta \to -\beta$,
implying $\gamma_*(-\beta) = \gamma_*(\beta)$.

At $\beta=0$, the critical attraction for stability of the GP functional is the constant
$$\gamma_*(0) =: \CLGN \approx 0.931 \times 2\pi.$$ 
It optimizes a quantitative, local uncertainty principle (due to Ladyzhenskaya, Gagliardo and Nirenberg; see \cite{Weinstein-83,AtaLunNgu-24}),
and is realized by 
``Townes' soliton'' 
\cite{ChiGarTow-64,Fibich-15}, 
the real and unique solution (mod scaling and translation)
at $E(0,-\CLGN,0)=0$.

In \cite{AtaLunNgu-24} it was shown that $\gamma_*(\beta) = 2\pi\beta$ for $\beta \ge 2$, while for intermediate $0 < \beta < 2$ it holds that $\gamma_*(\beta) > \max\{\CLGN,2\pi\beta\}$.
Further, although as a limit
$E(\beta,-2\pi\beta,0)=0$ for all $\beta \in \R$, 
zero-energy solutions only exist for 
$\beta = 2,4,6,\ldots$,
and we refer to these solutions as 
{\bf ``\emph{nonlinear} Landau levels'' (NLLs)}.
Namely, 
at the self-dual coupling $\gamma = -2\pi\beta$,
with $B=2\pi\beta\varrho$, $\beta>0$,
we again identify one of the spin components of the supersymmetric Pauli operator \eqref{eq:Pauli}-\eqref{eq:Pauli-susy}:
\begin{align}
    &\cE_{\beta,-2\pi\beta,0}[\phi]
	= \bigl\langle\phi \big|
	(-i\nabla + \bA)^2 - B \big |\phi\bigr\rangle. \label{eq:E}
    \end{align}
If $\textstyle\bA = \frac{1}{2}(\partial_2,-\partial_1)\psi$, 
where the superpotential $\psi$
solves $-\Delta\psi = 2B$, 
we can use the 
supersymmetric factorization 
    \begin{align}
    \cE_{\beta,-2\pi\beta,0}[\phi]
	= \int_{\R^2} \bigl|(\partial_1 - i\partial_2)(e^{-\psi/2}\phi)\bigr|^2 e^{\psi}.
    \label{eq:susy-trick}
\end{align}
Compare the standard, \emph{linear} Landau problem, where $B>0$ is a constant and where the 
zero-energy subspace, the
\emph{lowest} Landau level, consists of all normalizable functions on the form $\phi = \overline{f(z)} e^{\psi/2}$, where $f$ is analytic in $z = x_1+ix_2 \in \C$,
and $\psi(\bx) = -B\lvert\bx\rvert^2$.
In our problem, 
$\psi$ in \eqref{eq:susy-trick} solves the nonlinear and nonlocal, 
generalized Liouville equation $-\Delta \psi = 4\pi\beta |f|^2 e^\psi$, 
yielding the exact solutions \cite{AtaLunNgu-24}
\begin{equation}\label{eq:uPQ-solution}
    f \propto P'Q-PQ', \qquad
    \phi_{P,Q} := \sqrt{\frac{2}{\pi\beta}} \, \frac{\overline{P' Q - P Q'}}{|P|^2 + |Q|^2},
\end{equation}
where $P,Q$ are linearly independent and coprime complex polynomials with $n := \max(\deg P,\deg Q) = \beta/2$.
This complete set of regular minimizers includes all the self-dual soliton solutions found by Jackiw and Pi: initially, radial vortex states (``vortex rings'') with $P=z^n$ and $Q=1$ \cite{JacPi-90b}, and eventually, the more general rationals $P/Q$ \cite{JacPi-90a}.
Thus, along the critical line $\gamma=-2\pi\beta$, $\beta \in \R$, the sequence of points $\beta=2n$ marks a phase transition from strictly positive (broken supersymmetry, see Equations \eqref{eq:Pauli} and \eqref{eq:Pauli-susy}) to zero energy and increasing degree of supersymmetry, exhibited by NLL manifolds of real dimension $4n$;
compare Figures~\ref{fig:stability} and \ref{fig:numerics-att} (top).

\section{Macroscopic energy and density in a harmonic trap: numerics}
\label{sec:macro}

\subsection{Thomas--Fermi model}

On macroscopic scales (on the order of the trap) and for $\beta$ sufficiently large, 
we consider as our effective model a Thomas--Fermi (TF) energy functional:
\begin{equation}\label{eq:TF-func}
    \cETF_{G,V}[\tilde\varrho]
	:= \int_{\R^2} \left[
	G{\tilde\varrho(\bx)}^2 + V(\bx) \tilde\varrho(\bx) \right] \mathrm{d}\bx,
\end{equation}
where $\tilde\varrho(\bx)$ denotes an effective coarse-grained average spatial density of the gas,
subject to the normalization $\int \tilde\varrho = 1$.
The Thomas--Fermi constant $G$ is a parameter to be determined.
Thus, the minimizing density in the harmonic trap $V=\Vh$
is the standard TF profile,
$\rhoTF(\bx) = (2G)^{-1}(\lambda^{\rm TF} - |\bx|^2)_+$, 
supported on a radius 
$\sqrt{\lambda^{\rm TF}} \propto G^{1/4}$,
with the minimizing energy
$\ETF(G) = \frac{4}{3}\sqrt{G/\pi}$.
For comparison,
$G=2\pi N$ 
yields the asymptotic density and
energy per particle of $N$ spinless fermions in the harmonic trap.

\begin{figure}
	\centering
	\includegraphics[scale=0.35,clip,trim=1.3cm 0.5cm 1.3cm 0.2cm]{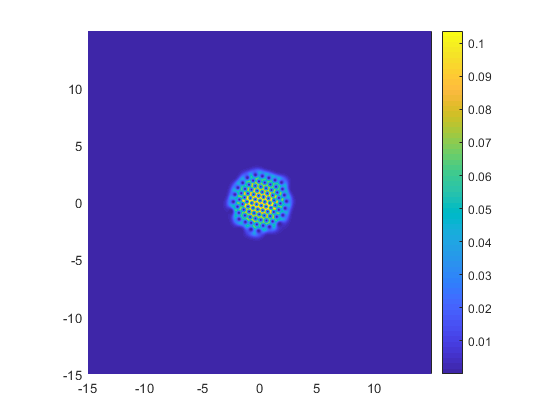}%
	\includegraphics[scale=0.35,clip,trim=1.3cm 0.5cm 1.3cm 0.2cm]{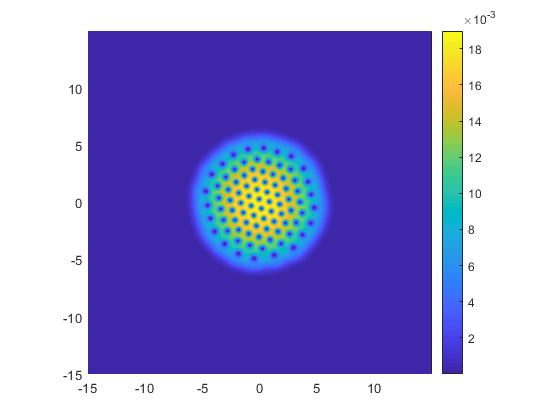}
	\caption{Vortex lattices in the densities of approximate minimizers for 
        $E(\beta=100,\gamma=-180\pi,\Vh) \approx 6.654$,
        respectively
        $E(\beta=100,\gamma=200\pi,\Vh) \approx 28.612$.}
	\label{fig:TF-comparison}
\end{figure}

Namely, 
motivated by a previous analytical \cite{CorLunRou-17,CorLunRou-18}
and numerical \cite{CorDubLunRou-19}
investigation of the low-energy states of the CSS functional \eqref{eq:CSS-func} for $\gamma=0$,
for $\beta \gg 1$ and $\gamma \propto \beta$ in the stable regime, 
we anticipate
the formation of a locally homogeneous 
vortex lattice on scales $\ell \sim \beta^{-1/4}$.
The intuition for this is that, writing
$\phi = \sqrt{\varrho} e^{i\varphi}$,
the magnetic-kinetic term splits as
\begin{equation}\label{eq:mag-kin-split}
    \bigl| (-i\nabla + \beta\bA_\varrho) \phi \bigr|^2 
    = |\nabla\sqrt{\varrho}|^2 
    + \bigl|\nabla\varphi + \beta\bA_\varrho\bigr|^2\varrho,
\end{equation}
where for locally homogeneous, radial density
(see \eqref{eq:mag-pot})
$$
    |\bA_\varrho(\bx)|^2 \sim \varrho(0)
$$
on macroscopic scales $|\bx| \sim 1$.
Therefore we see that,
unless a phase and a current $\nabla\varphi$ develops in $\phi$ to cancel 
some of this
self-magnetic potential, the resulting energy will be of relatively high order as $\beta \to \infty$.
For example, if the term \eqref{eq:mag-kin-split} remains on the order of $\beta^2 \varrho^2$ then the minimizing  density $\varrho$ spreads out on a macroscopic scale $|\bx| \sim \beta^{1/2}$ in the harmonic trap to yield a total energy on the order of $|\bx|^2 \sim \beta$.
However, it is possible to create unit vortices and corresponding unit flux domains on an average mesoscopic scale $\ell \sim \beta^{-1/4}$ to cancel the magnetic potential $\beta\bA_\varrho$ up to an error $\ell^{-2} \sim \beta^{1/2}$ on the average,
producing a magnetic-kinetic term on the order of $\beta\varrho^2$. 
Then the optimal density rather spreads out to a macroscopic scale $|\bx| \sim \ell^{-1} \sim \beta^{1/4}$ 
(covering roughly the area $\pi|\bx|^2 \sim \beta \pi\ell^2$)
to yield a total energy on the order of $|\bx|^2 \sim \ell^{-2} \sim \beta^{1/2}$.
Indeed this is what we observe to be the preferred solution,
with a locally homogeneous, triangular lattice of 
$O(\beta)$
singly quantized vortices (see Figure~\ref{fig:TF-comparison}).
However, in contrast to the Abrikosov lattices in strong external magnetic fields 
\cite{AftBlaDal-05,Fetter-09}, 
the lattice spacing here follows the average density profile sharply.

\begin{figure}
	\centering
	\begin{tikzpicture}
	\node[above right] at (0,5.85) {\includegraphics[scale=0.53,clip,trim=0.7cm 0cm 0.8cm 0.5cm]{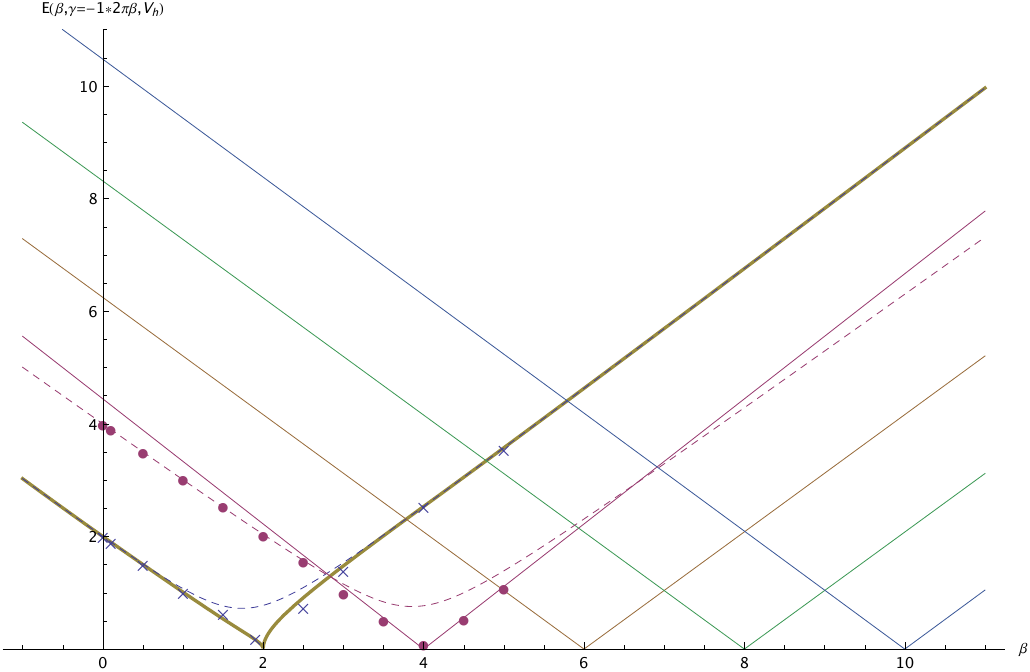}};
	\node[above right] at (8.3,5.85+0.25) {\scalebox{0.7}{$\beta$}};
	\node[above right] at (0.6,5.85+5.5) {\scalebox{0.7}{$E(\beta,\gamma=-2\pi\beta,\Vh)$}};
	\node[above right] at (0,0) {\includegraphics[scale=0.53,clip,trim=0.7cm 0cm 0.8cm 0.5cm]{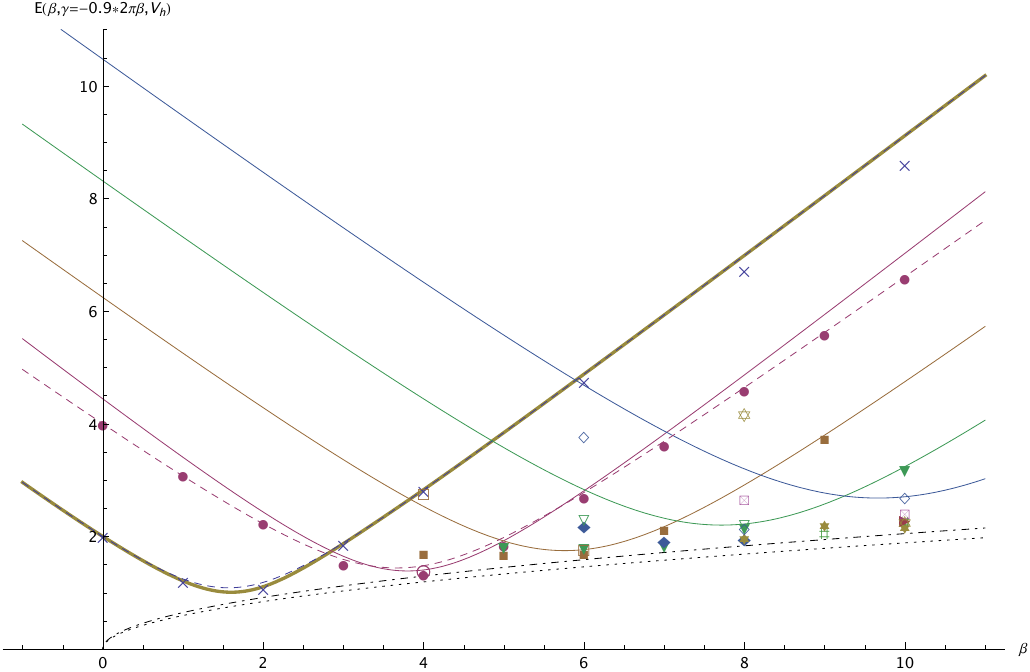}};
	\node[above right] at (8.3,0.25) {\scalebox{0.7}{$\beta$}};
	\node[above right] at (0.6,5.5) {\scalebox{0.7}{$E(\beta,\gamma=-0.9\times 2\pi\beta,\Vh)$}};
	\end{tikzpicture}	
	\caption{Analytical and numerical energies for attractive gas $\gamma = 2\pi\beta\kappa < 0$ at $\kappa = -1$ (``kreinyons'') resp. $\kappa = -0.9$.}
	\label{fig:numerics-att}
\end{figure}

Indeed, 
by general scaling arguments and local density approximation
on the macroscopic scale of the trap \cite{CorLunRou-17,CorLunRou-18}, we find
$$
    \bigl| (-i\nabla + \beta\bA_\varrho) \phi \bigr|^2
    \approx 2\pi c \beta \tilde\varrho^2,
$$
for a 
coarse-grained average density $\tilde\varrho(\bx) = (\pi r^2)^{-1} \int_{D(\bx,r)} \varrho$
over scales $r$ significantly larger than the vortex separation $\ell$ and significantly smaller than the trap.
Here $c$ denotes a universal constant corresponding to the homogeneous gas $\tilde\rho = \mathrm{const}$,
while $2\pi$ is the fermionic TF constant in our units, i.e. the energy per unit area and density of the homogeneous, spinless Fermi gas in the plane.
More generally, we expect
$$
    \bigl| (-i\nabla + \beta\bA_\varrho) \phi \bigr|^2
    + \gamma |\phi|^4
    \approx G \tilde\varrho^2,
$$
for some universal constant $G = G(\beta,\gamma)$ depending on both $\beta$ and $\gamma$,
and thus such an approximate TF 
description to hold: 
\begin{equation}\label{eq:TF-func-eff}
    \cE_{\beta,\gamma,V}[\phi] \ \approx \ 
	\cETF_{G,V}[\tilde\varrho].
\end{equation}
Therefore, for $V=\Vh$ we expect for the minimum energy 
$E(\beta,\gamma,\Vh) \simeq \ETF(G) = \frac{4}{3}\sqrt{G/\pi}$ 
and for the minimizer 
$\varrho(\bx) \approx \rhoTF(\bx)$
(in the coarse-grained sense)
if $\beta$ is large enough and $\gamma$ is of a similar order or smaller.

\begin{figure}
	\centering
	\begin{tikzpicture}
	\node[above right] at (0,5.85) {\includegraphics[scale=0.53,clip,trim=0.7cm 0cm 0.8cm 0.5cm]{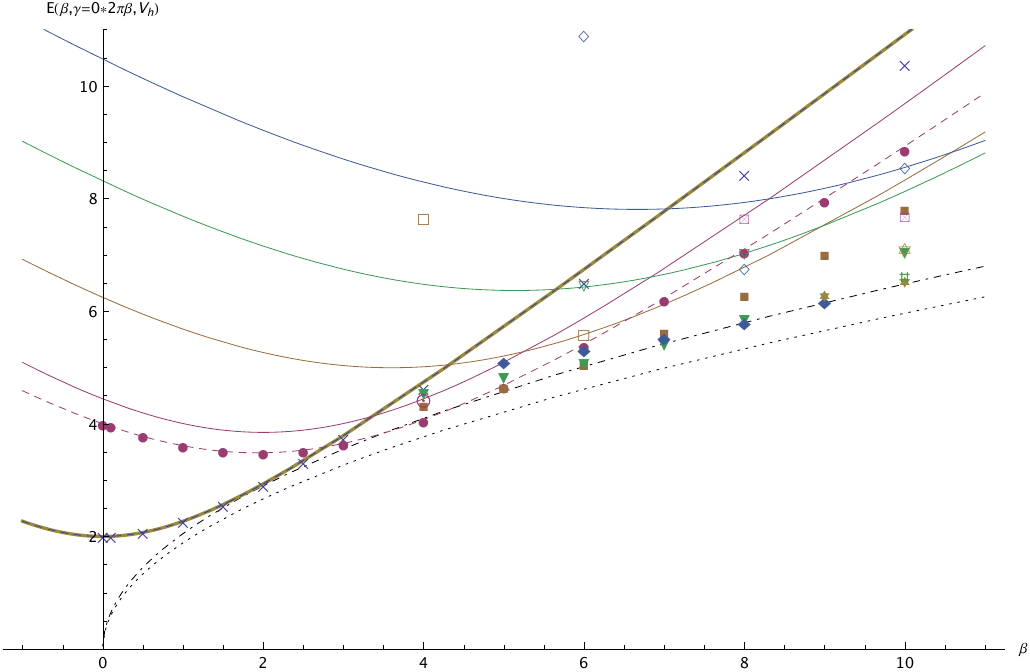}};
	\node[above right] at (8.3,5.85+0.25) {\scalebox{0.7}{$\beta$}};
	\node[above right] at (0.6,5.85+5.5) {\scalebox{0.7}{$E(\beta,\gamma=0,\Vh)$}};
	\node[above right] at (0,0) {\includegraphics[scale=0.53,clip,trim=0.7cm 0cm 0.8cm 0.5cm]{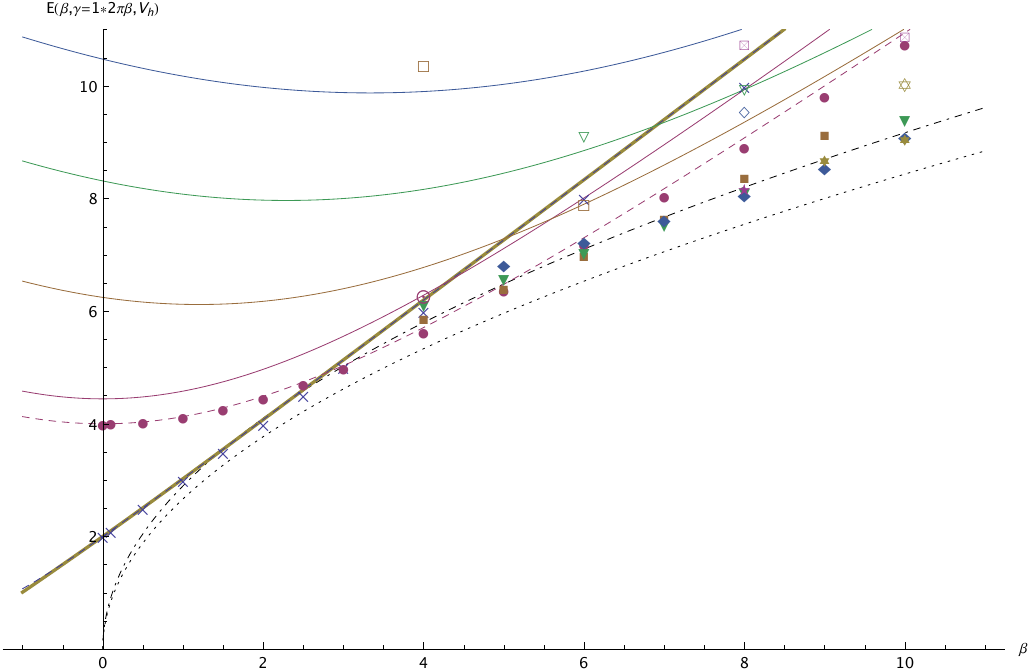}};
	\node[above right] at (8.3,0.25) {\scalebox{0.7}{$\beta$}};
	\node[above right] at (0.6,5.5) {\scalebox{0.7}{$E(\beta,\gamma=2\pi\beta,\Vh)$}};
	\node[above right] at (5.5,0.7) {\includegraphics[scale=0.53,clip,trim=0cm 0cm 0cm 0cm]{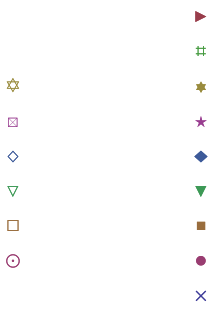}};
	\node[above right] at (5.5,2.95) {\scalebox{0.7}{NLL degree:}};
	\node[above right] at (5.9,0.74+1*0.32) {\scalebox{0.6}{1}};
	\node[above right] at (5.9,0.74+2*0.32) {\scalebox{0.6}{2}};
	\node[above right] at (5.9,0.74+3*0.32) {\scalebox{0.6}{3}};
	\node[above right] at (5.9,0.74+4*0.32) {\scalebox{0.6}{4}};
	\node[above right] at (5.9,0.74+5*0.32) {\scalebox{0.6}{5}};
	\node[above right] at (5.9,0.74+6*0.32) {\scalebox{0.6}{6}};
	\node[above right] at (7.2,3.63) {\scalebox{0.7}{\# vortices:}};
	\node[above right] at (7.6,0.74+0*0.32) {\scalebox{0.6}{0}};
	\node[above right] at (7.6,0.74+1*0.32) {\scalebox{0.6}{1}};
	\node[above right] at (7.6,0.74+2*0.32) {\scalebox{0.6}{2}};
	\node[above right] at (7.6,0.74+3*0.32) {\scalebox{0.6}{3}};
	\node[above right] at (7.6,0.74+4*0.32) {\scalebox{0.6}{4}};
	\node[above right] at (7.6,0.74+5*0.32) {\scalebox{0.6}{5}};
	\node[above right] at (7.6,0.74+6*0.32) {\scalebox{0.6}{6}};
	\node[above right] at (7.6,0.74+7*0.32) {\scalebox{0.6}{8}};
	\node[above right] at (7.6,0.74+8*0.32) {\scalebox{0.6}{9}};
	\end{tikzpicture}	
	\caption{Analytical and numerical energies for repulsive gas $\gamma = 2\pi\beta\kappa \ge 0$ at $\kappa=0$ resp. 1 (ideal anyons). See the text.}
	\label{fig:numerics-rep}
\end{figure}

\subsection{Numerical and analytical study of the spectrum}

In Figures~\ref{fig:numerics-att}--\ref{fig:numerics-rep} we study the actual energy spectrum 
of the CSS model \eqref{eq:CSS-func} along the line $\gamma=2\pi\beta\kappa$, 
for the range $\beta \in [0,10]$ and four different slopes $\kappa$.
When $\beta=\gamma=0$ it is simply the Schr\"odinger energy of the 2D harmonic
oscillator, with spectrum $2,4,6,\ldots$.
As upper bounds to the ground and excited states
we have computed the analytical energies of the first few radial vortex rings ($n=2,3,4,5$; optimally rescaled) and of an approximation to the ``versiera'' at $n=1$ with finite potential energy;
these are the solid curves in the plots;
see Appendix~\ref{app:analytic} for details.
We have also compared to the ground (Gaussian, at $n=1$) and first excited (single counter-rotating vortex, at $n=2$) states of the harmonic oscillator
(again optimally rescaled; dashed curves).

Further, general energies have been computed numerically (solid markers) by means of a
gradient-descent method, which is an extension of that used in \cite{CorDubLunRou-19} 
to the scalar self-interaction with parameter $\gamma$, 
with $(\beta,\gamma)$ in the stable regime.
See Appendix~\ref{app:numerics-matlab} for details and tabulated values, reaching up to $\beta = 100$.
We have tested the accuracy of the method at $\beta=0$ and various $\gamma$ to reproduce known GP energies; 
compare Table~\ref{tab:matlab-energies-0} and \cite[Table~3.5]{BaoCai-12}. 
Also, at $\gamma=0$ we reproduce the minimal energies of \cite{CorDubLunRou-19},
i.e.\ the minima follow closely the estimate $\ETF(2\pi c\beta)$ where 
the numerical fit
$c = 2\sqrt{\pi}/3 \approx 1.18 > 1$ is a renormalizing ``Abrikosov factor'' originating from the small-scale variations of the emerging vortex lattice.
Interestingly, we may observe that this behavior holds also for small $\beta$, 
all the way down to $\beta \gtrsim 2$.
The critical $\beta$ for nucleation of a single vortex is just slightly below 3.

For a general estimate of the ground-state energy at $\gamma=2\pi\beta\kappa$, $\kappa \ge -1$, 
we have compared with the linear interpolation 
$G = 2\pi c(1+\kappa)\beta$, i.e.
$\ETF(G) 
= \ETF(2\pi\beta)\sqrt{c}\sqrt{1+\kappa}$
(dash-dotted curve), which provides a fit to numerics to within a few percent for $\beta \gtrsim 3$ and $-0.9 \lesssim \kappa \lesssim 2$.
In the early work \cite{ChiSen-92} there was an estimate given for the corresponding energy for ideal anyons at $\kappa=1$ which is ca 7\% smaller.
The radius of the TF profile seen in Figure~\ref{fig:TF-comparison} is also accounted for by the estimate $G^{1/4} \propto (1+\kappa)^{1/4}\beta^{1/4}$.
An analytical absolute lower bound for the energy (assuming regularity and singlevaluedness of $\phi$) is provided by the minimizing quotient
$\gamma_*(\beta) \to 2\pi\beta$ 
for $\beta \ge 2$
and thus differs from our estimate by a factor $\sqrt{c}$, 
i.e.\ 9\% (dotted curve).
Note that for ideal anyons that is still $\sqrt{2}-1 \approx 41\%$ higher than the ideal Fermi energy.
Based on the numerical data and comparing with the exact result \eqref{eq:true-gs-energy},
we also conjecture that the real-valued minimizer at $\kappa = \pm 1$ has a linearly interpolating energy
$|2\pm\beta|$.
This, as well as other linear features of the $\kappa=\pm 1$ spectra, resembles the known linear/noninterpolating exact states of the $N$-anyon spectrum \cite{Chou-91a,MurLawBhaDat-92,ChiSen-92,AtaGirLun-25}.

Lastly,
for $\beta = 2n \in \{4,6,8,10\}$ 
we have studied how well the NLLs \eqref{eq:uPQ-solution} approximate the CSS energy spectrum for subcritical coupling $\kappa > -1$.
Namely, we have found analytical expressions for these NLLs
by solving the inverse Wron\-sk\-ian problem $P,Q \mapsto W(P,Q) := P'Q-PQ'$; see Appendix~\ref{app:wronskian}.
Then we used numerical integration together with optimization of the placement of zeros/vortices 
to estimate their minimal energy
\begin{equation}\label{eq:NLL-energy}
    \cE_{\beta,\gamma,\Vh}[\phi_{P,Q}] 
    = \int_{\R^2} \left[
		(4\pi n + \gamma) |\phi_{P,Q}|^4
		+ \Vh |\phi_{P,Q}|^2 \right].
\end{equation}
These energies are given the hollow markers in the plots; 
see Appendix~\ref{app:numerics-mathematica} 
for details and tabulated values.
Although for small $\beta$ it is very costly to break the radial symmetry, there is a trend that for larger $\beta$ and smaller $\gamma$ this cost can be reduced in sufficiently symmetric vortex configurations. Indeed, for $\beta=8$, $\kappa=-0.9$ an NLL state with 4 separated vortices has lower energy than any of the radial vortex rings (see Fig.~\ref{fig:numerics-att}), while for $\beta=10$ the NLLs for 5 and 6 separated vortices yield significantly better energies even for positive $\gamma$ (see Fig.~\ref{fig:numerics-rep}).

\section{Conclusions and outlooks}
\label{sec:outlooks}

We have derived and analyzed an effective CSS model \eqref{eq:CSS-func} for the interacting many-anyon gas in the almost-bosonic limit, which manifests a spectrum of energies/states that are well describable using a family of exact NLLs \eqref{eq:uPQ-solution} as well as an effective TF model \eqref{eq:TF-func} on macroscopic scales.
Although we have here focused on small flux $\beta \lesssim 10$, the Matlab part of our numerics extends to ca $\beta \lesssim 140$ 
(see Table~\ref{tab:matlab-energies-50} in Appendix~\ref{app:numerics-matlab}).
Our study of exact NLLs is mainly limited by the increasing complexity of the inverse Wronskian problem, however we may extend relatively easily to some further symmetry-reduced cases for 
$\beta \ge 10$
(see \cite{Getzner-26}).
Ongoing work also includes external magnetic fields 
(see \cite{Scribano-26})
and the Abrikosov problem for homogeneous self-generated vortex lattices.
Further,
in our analysis of NLLs we have assumed regularity and singlevaluedness of CSS wave functions $\phi$. 
We believe that in order to obtain even lower energies, 
to eventually match the ideal Fermi energy at $\alpha \to 1$, we must relax these assumptions. This also leads to the study of systems of CSS functions and Liouville--Toda-type equations (cf.~\cite{DunJacPiTru-91,Dunne-92,Kim-etal-93}), which will be addressed in future work.
Some new insight into the fermionic side of the problem, via fermionic Hartree and magnetic TF theory, has been recently discussed in \cite{LevLunRou-25} (see also \cite{Hu-etal-21}).
In fact this shows that, as we approach fermions, the interplay of different self-generated Landau levels will become increasingly relevant, and thus the CSS functional considered here is more appropriately replaced by that for a system of CSS wave functions.

We hope that our work can help make progress on the understanding of various universal aspects of anyon gases, such as superconductivity (see \cite{CheWilWitHal-89,Wilczek-90,Lundholm-23}, and e.g. the recent experiments \cite{Xu-etal-25}). Indeed, our stability analysis of the mesoscopic model indicates that collective resonances form at specific values of the parameters, defined by the NLLs or Jackiw--Pi solitons. However, more work is required to nail down the specifics, including an attractive spin-orbit coupling in the microscopic model 
(see \cite{AtaGirLun-25}). 

In conclusion, we believe that important keys to understanding the complex interpolation between the almost-bosonic and almost-fermionic anyon spectra will come via the study of NLLs and their possible extension to more general functions.
Also, exploring the two-parameter energy landscape may uncover fruitful and highly nontrivial relations between the CSS, GP, TF, and other models.
For example, despite our assumption in this work on the individual anyonic flux being small, we may expect some of the studied features to extend more generally due to the universal nature of the CSS and TF models.

\begin{ack}
We are grateful to Romain Duboscq for the original Matlab code upon which much of our numerical work is based.
We also thank the anonymous referees for their comments and questions which have helped to improve the manuscript.
A.A., A.E., F.G. and D.L.~thank
the Department of Mathematics of Politecnico di Milano for its kind hospitality 
during the spring of 2025 intensive period ``Quantum Mathematics at Polimi''.
D.L.~also thanks the GSSI L'Aquila for its hospitality.
All of us have benefited from discussions at the Institut Mittag-Leffler workshop ``Anyons from small to large scales'' in July 2025.
Financial support from the Swedish Research Council 
(D.L., grant no.\ 2021-05328, ``Mathematics of anyons and intermediate quantum statistics'') 
is gratefully acknowledged. A.A. thanks Kaj Nystr\"om for the financial support for traveling to the relevant conferences. 
The author T.G.~acknowledges the Gran Sasso Science Institute for the financial support making this work possible.
A.E.~gratefully thanks the Center for Interdisciplinary Mathematics at Uppsala University
as well as C.F.~Liljewalchs stipendiestiftelse for their financial support. D.-T.N. was funded by the Postdoctoral Scholarship Programme of Vingroup Innovation Foundation (VINIF), VinUniversity, code VINIF.2025.STS.15, and also gratefully acknowledges the support of the Knut and Alice Wallenberg Foundation through the grant of Prof. Andreas Str\"ombergsson during his postdoctoral period at Uppsala University (2022–2024).
\end{ack}


\appendix
\section{}
\label{app:analytic}

In these appendices we give all the information necessary to reproduce the plots
of Figs.~\ref{fig:numerics-att}--\ref{fig:numerics-rep}.
After setting our conventions,
we first derive some explicit analytical NLL states 
\eqref{eq:uPQ-solution}
by solving the inverse
Wronskian problem.
Then we analytically compute the energies of some related trial states,
including a reference Gaussian, 
the first excited harmonic oscillator
energy eigenstates,
powers of the ``versiera'' ansatz, 
as well as 
NLL ``vortex rings''.
Finally, we give tables of numerical data.

\subsection{Notation and choice of gauge}
\label{app:notation}

The $R$-extended anyon gauge vector potential is defined by $\bA^R_j := \sum_{k \neq j} \frac{(\bx_j-\bx_k)^\perp}{|\bx_j-\bx_k|_R^2}$ where $\bx^{\perp} := (-x_2,x_1)$ denotes the counterclockwise $90^\circ$ rotation for $\bx = (x_1,x_2) \in \R^2$, and
$|\bx|_R := \max\{R,|\bx|\}$.
The effective gauge potential for a given density distribution $\varrho \in L^1(\R^2;\R_+)$ is defined by the convolution
\begin{align*}
  \bA_\varrho(\bx) :=  \int_{\R^2} \frac{(\bx-\by)^{\perp}}{|\bx-\by|^2} \varrho(\by) \, \mathrm{d} \by.
\end{align*}
In both vector potentials we have chosen the Coulomb gauge $\nabla \cdot \bA = 0$, and they generate the magnetic fields
$$
    B^R_j = \curl_{\bx_j} \bA^R_j = \begin{cases}
    \frac{2}{R^2} \sum_{k \neq j} \1_{D(\bx_k,R)}(\bx_j), & R>0, \\
    2\pi \sum_{k \neq j} \delta(\bx_j - \bx_k), & R=0,
    \end{cases}
$$
respectively
$$
    B_\varrho = \curl \bA_\varrho = 2\pi \varrho
$$
(in two dimensions, the magnetic field 
$B = \partial_1 A_2 - \partial_2 A_1$
is a (pseudo-)scalar quantity),
where we used that $\Delta \log |\bx| = 2\pi\delta(\bx)$.
Throughout, we use the notation $D(\bx,r)$ for a disk of radius $r$ at $\bx$,
and $\1_\Omega$ for the characteristic function or projection operator supported on $\Omega$.
Also, $x_+ := \max\{0,x\}$.
Sometimes we use complex notation for the coordinates
$z = x_1 + ix_2 = re^{i\theta} \in \C$, 
$\bar{z} = x_1 - ix_2$.

Note that,
because of our gauge choice and
vanishing condition 
$\lim_{\bx \to \infty} \bA_\varrho = 0$,
for a radially symmetric density $\varrho(\bx) = \varrho(r)$, we must have
\begin{equation}
    \bA_\varrho(\bx) = A(r)\be_\theta,
\end{equation}
for some radial function $A(r)$. Stokes' theorem then implies that
\begin{equation*}
    2\pi rA(r) = \oint_{\partial D(0,r)} \bA_{\varrho} \cdot \mathrm{d}\mathbf{s} = \int\limits_{D(0,r)} \curl \bA_\varrho = 2\pi \int\limits_{D(0,r)} \varrho,
\end{equation*}
so
\begin{equation}
    A(r) = \frac{1}{r} \int\limits_{D(0,r)} \varrho = \frac{2\pi}{r} \int_0^r \varrho(s)s \, \mathrm{d}s.
    \label{eq:mag-pot}
\end{equation}

\subsection{NLL states and inverse Wronskian problem}
\label{app:wronskian}

The general formula \eqref{eq:uPQ-solution} for NLL states is parametrized by the space of solutions $(P,Q)$ of the inverse Wronskian problem
\begin{equation} \label{eq:Wronskian-problem}
    W(P,Q) := P'Q-PQ' = f,
\end{equation}
where $f$ is a given polynomial
and $P$ and $Q$ are coprime 
(and linearly independent)
polynomials satisfying the degree condition $\max\{\deg P,\deg Q\} = n = \beta/2 > 0$.

Due to the determinantal property of the Wronskian and the symmetry of the amplitude $|P|^2+|Q|^2$,
the corresponding NLL state $\phi_{P,Q}$ given by \eqref{eq:uPQ-solution} is invariant under pointwise action of the group $\mathrm{SU}(2)\times\R^+$ on the space of polynomial pairs $(P,Q)$. It follows that we may without loss of generality 
assume that either $P$ or $Q$ is monic, 
and, e.g., that $\deg Q < \deg P = n = \beta/2$. 
We may also simplify the problem \eqref{eq:Wronskian-problem} by adding any multiple of $Q$ to $P$.
If we identify states that differ only by multiplication by a constant phase factor $e^{i\theta}$, we have a further $\sU(1)$-symmetry, for a full $\sU(2)\times\R^+$-symmetry in total.
See \cite{AtaLunNgu-24} for further details.

Thus, given the location of vortices/zeros $w_k \in \C$, i.e. $f(z) \propto \prod_k (z-w_k)$,
we may enumerate all NLL solutions with these vortices by
\[
    (\tilde{P},\tilde{Q}) = (P+\eta Q,\lambda^{-1} Q), \qquad \eta\in\C, \quad \lambda > 0,
\]
and
\begin{equation} \label{eq:uPQ-state}
    \phi_{\tilde{P},\tilde{Q}} = \frac{1}{\sqrt{\pi n}} \frac{\overline{W(P,Q)}}{\lambda\lvert P + \eta Q \rvert^2 + \lambda^{-1} \lvert Q \rvert^2},
\end{equation}
where $(P,Q)$ are particular solutions to the Wronskian problem \eqref{eq:Wronskian-problem} of the simplified form.

Notice that the number of vortices for a given NLL state $\phi_{P,Q}$ (counting multiplicities) is precisely $\deg W(P,Q)$, which for fixed $\beta/2 = n = \max\{\deg P, \deg Q\}$
is restricted by the inequalities
\cite{AtaLunNgu-24}
\begin{equation} \label{eq:wronskian-deg-range}
    n-1 \leq \deg W(P,Q) \leq 2n-2.
\end{equation}
Hence the $n$:th NLL space contains states with between $n-1$ and $2n-2$ vortices. We will now list some known NLL solutions for various $n$.

For any positive integer $n = \beta/2 = 1,2,3,\ldots$, we may consider the degenerate case
\[
    f(z) = n(z-w)^{n-1}, \qquad w\in\C.
\]
A particular solution to the inverse Wronskian problem $W(P,Q) = f$ is then
\begin{equation} \label{eq:wronsk-vortex-ring}
    P(z) = (z-w)^n, \qquad Q(z) = 1.
\end{equation}
The corresponding state on the form \eqref{eq:uPQ-state} is the ``\textbf{vortex ring}''
\begin{equation} \label{eq:vortex-ring-state}
    \phi_{\tilde{P},\tilde{Q}}(z) = \sqrt{\frac{n}{\pi}} \frac{(\bar{z}-\bar{w})^{n-1}}{\lambda\lvert (z-w)^n + \eta \rvert^2 + \lambda^{-1}},
\end{equation}
for $\lambda > 0$ and $\eta\in\C$.
For $\eta=0$, the corresponding density is radially symmetric about $w$.
These solutions (for $w=\eta=0$) were first described by Jackiw and Pi in \cite{JacPi-90b}.
For $n>1$, they have a vortex (zero) of degree $n-1$ at $z = w$, but in the special case $n=1$, we get the real-valued and positive ``\textbf{versiera}'' solution
\begin{equation} \label{eq:versiera-state}
    \phi(z) = \frac{1}{\sqrt{\pi}} \frac{1}{\lambda \lvert z-w \rvert^2 + \lambda^{-1}},
\end{equation}
which lacks vortices (note that the parameter $\eta$ can be absorbed into $w$ in this case). The family of versiera solutions, parametrized by the scale $\lambda>0$, center $w \in \C$, and a constant phase, is the complete manifold of NLL states for $n = 1$.

    For fixed $n = \beta/2$, the $4n$-dimensional manifold of NLL states foliates over the degree $d = \deg f$, $n-1 \leq d \leq 2(n-1)$. For $n = 1$, we must have $d = 0$, so the versiera solutions are the only possible solutions. For $n=2$ and $d = 1$, we have the particular solution
    \begin{equation} \label{eq:wronsk-beta4-deg1}
        f(z) = z-w, \quad 
        P(z) = z^2-2wz, \quad 
        Q(z) = \frac{1}{2},
    \end{equation}
    corresponding to the family of vortex ring solutions \eqref{eq:vortex-ring-state} of degree 1, and for $d = 2$ we have the particular solution
    \begin{align}
        f(z) &= (z-w_1)(z-w_2), \nonumber\\
        P(z) &= z^2 - w_1w_2, \label{eq:wronsk-beta4-deg2}\\
        Q(z) &= z - \frac{w_1+w_2}{2}. \nonumber
    \end{align}
    We must have $w_1 \neq w_2$, for otherwise $P$ and $Q$ are not coprime. Hence, these states have two separated vortices located at $w_1$ and $w_2$. 
    With the additional degrees of freedom in \eqref{eq:uPQ-state},
    these two families of solutions exhaust the NLL manifold for $n=2$ (this was discussed in \cite[Section 2.3]{AtaLunNgu-24}).

    For $n=3$, we have $2\leq d \leq 4$ by \eqref{eq:wronskian-deg-range}. For $d = 2$, we have the particular solution
    \begin{align}
        f(z) = (z-w_1)(z-w_2), \nonumber\\
        P(z) = z^3 - 3\frac{w_1+w_2}{2}z^2 +3 w_1w_2z, && Q(z) = \frac{1}{3}, \label{eq:wronsk-beta6-deg2}
    \end{align}
    with $w_1,w_2\in\C$ arbitrary. If $w_1=w_2$, this is a vortex ring \eqref{eq:vortex-ring-state} of degree 2. For $d=3$, we have two 
    particular solutions:
    \begin{align}
        f(z) &= (z-w_1)(z-w_2)(z-w_3), \nonumber\\
        P(z) &= z^3 - A_\pm z^2 + 2s_3, \label{eq:wronsk-beta6-deg3}\\
        Q(z) &=\frac{1}{2} z - \frac{1}{6}A_\mp, \nonumber
    \end{align}
    where
    \begin{align*}
         A_\pm &:= s_1 \pm \sqrt{s_1^2-3s_2}.
    \end{align*}
    Here
    \begin{equation*}
        s_k = s_k(w_1,\ldots,w_d) := \sum_{1 \leq i_1 < \cdots < i_k \leq d} w_{i_1} \cdots w_{i_k}
    \end{equation*}
    is the $k$:th elementary symmetric polynomial in the variables $w_1,\ldots,w_d$. In particular,
    \begin{align*}
        s_1(w_1,w_2,w_3) &= w_1+w_2+w_3, \\
        s_2(w_1,w_2,w_3) &= w_1w_2+w_1w_3+w_2w_3, \\
        s_3(w_1,w_2,w_3) &= w_1w_2w_3.
    \end{align*}
    For $d = 4$, we have the two particular solutions
    \begin{align}
        f(z) &= (z-w_1)(z-w_2)(z-w_3)(z-w_4), \nonumber\\
        P(z) &= z^3 - \frac{1}{2}B_\mp z + \frac{1}{2}s_3, \label{eq:wronsk-beta6-deg4}\\
        Q(z) &= z^2 - \frac{1}{2}s_1z + \frac{1}{6}B_\pm, \nonumber
    \end{align}
    where
    \begin{align*}
        B_\pm &:= s_2 \pm \sqrt{s_2^2 - 3s_1s_3 + 12s_4}.
    \end{align*}
    The above particular solutions yield all NLL states for $n=3$ via \eqref{eq:uPQ-state},
    and $P,Q$ are coprime at least if all the roots $w_j$ are disjoint 
    (consider $f'=P''Q-PQ''$).

    For $n=4$ and $n=5$, we will be content with giving some special solutions with additional symmetries that have been used in the numerics (see Appendix~\ref{app:numerics-mathematica}). With $d = \deg f$ as usual, we have $3\leq d \leq 6$, respectively $4 \leq d \leq 8$, by \eqref{eq:wronskian-deg-range}. 
    For $n=4$ and $d=3,4$, we place the vortices at a distance $R$ from the origin along the roots of unity, $w_k = Re^{i2\pi k/d}$. This yields the particular solutions
    \begin{equation}\label{eq:wronsk-beta8-deg3}
        f(z) = z^3 - R^3, \quad
        P(z) = z^4 - 4 R^3 z, \quad
        Q(z) = \frac{1}{4},
    \end{equation}
    for $d = 3$, and
    \begin{equation}\label{eq:wronsk-beta8-deg4}
        f(z) = z^4 - R^4, \quad
        P(z) = z^4 +3 R^4, \quad
        Q(z) =\frac{1}{3} z,
    \end{equation}
    for $d = 4$. For $d = 5$, we place one vortex at the origin and the other four along the roots of unity, yielding the particular solution
    \begin{equation}\label{eq:wronsk-beta8-deg5}
        f(z) = z(z^4-R^4), \quad
        P(z) =z^4+R^4, \quad
        Q(z) = \frac{1}{2} z^2.
    \end{equation}
    For $d=3$ we can allow $R \ge 0$, while for $d=4,5$
    we require $R>0$ for coprimality.
    For $n=4$ and $d=6$ no sufficiently symmetric solutions and densities have been found (see \cite{Getzner-26} for examples).

    For $n=5$ and $d=4,5$ we place all vortices along roots of unity at distance $R$, yielding the particular solutions
    \begin{equation}\label{eq:wronsk-beta10-deg4}
        f(z) = z^4 - R^4, \quad
        P(z) = z^5 - 5 R^4 z, \quad
        Q(z) = \frac{1}{5},
    \end{equation}
    for $d = 4$, and
    \begin{equation}\label{eq:wronsk-beta10-deg5}
        f(z) = z^5 - R^5, \quad
        P(z) = z^5 + 4 R^5, \quad
        Q(z) =\frac{1}{4} z,
    \end{equation}
    for $d = 5$. For $d = 6$, we place one vortex at the origin and the other five along the roots of unity, yielding the particular solution
    \begin{equation}\label{eq:wronsk-beta10-deg6}
        f(z) = z(z^5-R^5), \quad
        P(z) = z^5+\frac{3}{2}R^5, \quad
        Q(z) = \frac{1}{3} z^2.
    \end{equation}
    For $d=4$ we can allow $R \ge 0$, while for $d=5,6$ we require $R>0$ for coprimality.

\subsection{Analytical energies of trial states}
\label{app:trial}

We will now calculate the exact energy $\cE[\phi] = \cE_{\beta,\gamma,V}[\phi]$ analytically for certain trial CSS states $\phi$ in the harmonic trap $V(z) = \Vh(z) = \lvert z \rvert^2$. The parameters $\beta$ and $\gamma$ are considered fixed.

First we make some general observations. By expanding the kinetic-magnetic term $\lvert (\nabla + i\beta\bA_\varrho) \phi \rvert^2$, the energy functional can be recast into the form
\begin{equation}
    \cE[\phi] = \int_{\R^2} \left[ \lvert \nabla \phi \rvert^2 + 2 \beta \bA_\varrho \cdot \bJ_{\phi} + \beta^2 \lvert \bA_\varrho \rvert^2 \varrho + \gamma \varrho^2 + V\varrho \right],
\end{equation}
where
\begin{equation}
    \bJ_\phi := \frac{i}{2}\left(\phi\nabla\overline{\phi} - \overline{\phi}\nabla\phi\right)
\end{equation}
is the probability current induced by $\phi$. When $\phi$ has a constant phase, $\bJ_\phi = 0$.

Now we turn to the question of finding the optimal length scale in the harmonic potential $\Vh$, or equivalently the optimal scale of the trial state. If $\phi$ is a normalized trial state and $\lambda > 0$, let
\begin{equation}
    \phi_\lambda(z) := \lambda \phi (\lambda z).
\end{equation}
The rescaling $\phi \to \phi_\lambda$ induces rescalings of the quantities derived from $\phi$:
\begin{align*}
    \nabla\phi_\lambda(z) &= \lambda^2 (\nabla \phi) (\lambda z); \\
    \varrho_\lambda(z) &:= \lvert \phi_\lambda(z) \rvert^2 = \lambda^2 \varrho(\lambda z); \\
    \bA_\lambda(z) &:= \bA_{\varrho_\lambda}(z)
    = \lambda \bA_\varrho(\lambda z); \\
    \bJ_\lambda(z) &:= \bJ_{\phi_\lambda}(z) 
    = \lambda^3 \bJ_\phi(\lambda z).
\end{align*}
As a result, if we let
\begin{align}
    K_\lambda &:= \int_{\R^2} \left[ \lvert (\nabla + i\beta\bA_\lambda)\phi_\lambda \rvert^2 + \gamma \varrho_\lambda^2 \right];
    \label{eq:Klambda}\\
    U_\lambda &:= \int_{\R^2} \Vh\varrho_\lambda,
    \label{eq:Ulambda}
\end{align}
then $K_\lambda = \lambda^2 K_1$ and $U_\lambda = \lambda^{-2}U_1$,
while $\int_{\R^2} \varrho_\lambda = 1$.
Hence, the total energy is
\begin{equation} \label{eq:total-energy}
    E_\lambda := \cE[\phi_\lambda] = K_\lambda + U_\lambda = \lambda^2 K_1 + \lambda^{-2} U_1,
\end{equation}
which achieves its minimum at the optimal scaling parameter
\begin{equation} \label{eq:optimal-scaling}
    \lambda_* := (U_1/K_1)^{1/4}.
\end{equation}
Using \eqref{eq:total-energy}, the corresponding minimal energy is
\begin{equation} \label{eq:optimal-energy}
    E_{\lambda_*} = 2\sqrt{K_1 U_1}.
\end{equation}

\subsection{Gaussian}
\label{app:gaussian}

We now compute the energy $\cE[\phi] = \cE_{\beta,\gamma,\Vh}[\phi]$ analytically for some trial states $\phi$, starting with a radially symmetric, Gaussian state
\begin{equation}
    \phi(z) = \frac{1}{\sqrt{\pi}}e^{-r^2/2},
\end{equation}
where $r = \lvert z \rvert$.
The corresponding density is
\[
    \varrho(z) = \lvert \phi(z) \rvert^2 = \frac{1}{\pi}e^{-r^2},
\]
satisfying the normalization condition $\int_{\R^2} \varrho = 1$.
Using \eqref{eq:mag-pot}, we get the corresponding magnetic potential
\begin{equation}
    \bA_\varrho(z) = \frac{2}{r}\left(\int_0^r s e^{-s^2} \, ds\right) \be_\theta = \frac{1}{r} (1-e^{-r^2}) \be_\theta.
\end{equation}
Since $\phi$ is real, $\bJ_\phi = 0$.
The kinetic energy is
\[
    \int_{\R^2} \lvert \nabla \phi \rvert^2 = 2\pi \int_0^\infty \left\lvert \frac{\partial \phi}{\partial r} \right\rvert^2 r \, \mathrm{d}r = 2 \int_0^\infty r^3 e^{-r^2} \, \mathrm{d}r = 1,
\]
where we used the standard integral
\begin{equation} \label{eq:std-int-xex2}
    \int_0^\infty x^{2k+1}e^{-ax^2} \, \mathrm{d}x = \frac{k!}{2a^{k+1}},
    \qquad k \in \N_0,\ a>0.
\end{equation}
The magnetic self-energy is given by
\[
    \int_{\R^2} \lvert \bA_\varrho \rvert^2 \varrho = 2\int_0^\infty \frac{(1-e^{-r^2})^2e^{-r^2}}{r} \, \mathrm{d}r = \log\frac{4}{3},
\]
which can be derived from the standard integral
\begin{equation}
    \int_0^\infty \frac{e^{-ax} - e^{-bx}}{x} \, \mathrm{d}x = \log\frac{b}{a}, \qquad a,b > 0,
\end{equation}
by expanding the integrand and substituting $t=r^2$.
The scalar self-interaction energy is
\begin{align*}
    \int_{\R^2} \varrho^2 = \frac{2}{\pi} \int_0^\infty re^{-2r^2} \, \mathrm{d}r = \frac{1}{2\pi},
\end{align*} and thus
$$
    K_1 = 1 + \beta^2\log\frac{4}{3} + \frac{\gamma}{2\pi}.
$$
Finally, the potential energy in the harmonic trap $V(z) = \Vh(r) = r^2$ is
\[
   U_1= \int_{\R^2} \Vh\varrho = 2 \int_0^\infty r^3 e^{-r^2} \, \mathrm{d}r = 1,
\]
once again using \eqref{eq:std-int-xex2}.

Now consider $\phi_\lambda$.
At the optimal scaling parameter $\lambda=\lambda_*$ (see \eqref{eq:optimal-energy}), the total energy is
\begin{equation}
    \cE_{\beta,\gamma,\Vh}[\phi_{\lambda_*}] = 2\sqrt{1 + \beta^2\log\frac{4}{3} + \frac{\gamma}{2\pi}}.
\end{equation}
    With $\gamma=2\pi\beta\kappa$, this is shown as the dashed blue curve in the plots.

\subsection{Excited harmonic oscillator eigenstates}
\label{app:harmosc}

The Gaussian is the ground state of the harmonic oscillator. Now, we consider the two angular momentum eigenmodes of the first excited state:
\[
    \phi^+(z) = \frac{1}{\sqrt{\pi}}ze^{-\lvert z \rvert^2/2}, \qquad 
    \phi^-(z) = \frac{1}{\sqrt{\pi}}\bar{z}e^{-\lvert z \rvert^2/2}.
\]
In polar coordinates $z = re^{i\theta}$, we have
\[
    \phi^\pm(z) = \frac{1}{\sqrt{\pi}}re^{-r^2/2}e^{\pm i\theta}.
\]
The density is the same for both modes:
\[
    \varrho(z) = \lvert \phi^\pm(z) \rvert^2 = \frac{1}{\pi}r^2e^{-r^2},
\]
and this is normalized to $\int_{\R^2} \varrho = 1$.
Using \eqref{eq:mag-pot},
the corresponding magnetic potential is $\bA_\varrho = A(r)\be_\theta$, with
\begin{align*}
    A(r) = \frac{2}{r} \int_0^r s^3e^{-s^2} \, \mathrm{d}s = \frac{1-(1+r^2)e^{-r^2}}{r}.
\end{align*}
Differentiating $\phi^\pm$ yields
\begin{align*}
    \nabla \phi^\pm &= \frac{1}{\sqrt{\pi}}e^{\pm i\theta}e^{-r^2/2} \left[(1-r^2)\be_r \pm i\be_\theta \right] \\
    &= \frac{\phi_\pm}{r}\left[(1-r^2) \be_r \pm i\be_\theta \right].
\end{align*}
The currents for the two modes are therefore
\[
    \bJ_\pm := \bJ_{\phi^\pm} = \pm \frac{\varrho}{r}\be_\theta = \pm \frac{1}{\pi}re^{-r^2}\be_\theta.
\]
The harmonic oscillator energy contributions are
\[
    \int_{\R^2} \lvert\nabla\phi^\pm\rvert^2 = \int_{\R^2} \Vh\varrho = 2,
\]
which can be verified either by direct computation using \eqref{eq:std-int-xex2}, or by using that $\phi^\pm$ are energy eigenstates of the harmonic oscillator.
The scalar self-energy is given by
\[
    \int_{\R^2} \varrho^2 = \frac{2}{\pi}\int_0^\infty r^5e^{-2r^2} \, \mathrm{d}r = \frac{1}{4\pi}.
\]
The magnetic self-energy common to both modes is given by
\[
    \int_{\R^2} \lvert \bA_\varrho \rvert^2 \varrho = 2 \int_0^\infty re^{-r^2} \left( 1 - 2(1+r^2)e^{-r^2} \right) \, \mathrm{d}r = \frac{7}{54},
\]
again using \eqref{eq:std-int-xex2}.
The contribution due to the current is
    \[
        \int_{\R^2} \bJ_\pm \cdot \bA_\varrho = \pm 2\int_0^\infty re^{-r^2}\left(1-(1+r^2)e^{-r^2}\right) \, \mathrm{d}r = \pm \frac{1}{4}.
    \]
The total energy is thus
$$
    K_1 = 2 + \frac{1}{4\pi}\gamma + \frac{7}{54}\beta^2 \pm \frac{1}{2}\beta,
    \qquad
    U_1 = 2.
$$

At the optimal scaling parameters $\lambda = \lambda_\pm$ (see \eqref{eq:optimal-energy}), we have the minimal energies
\begin{equation}
    \cE_{\beta,\gamma,\Vh}\left[\phi^\pm_{\lambda_\pm}\right] = 2\sqrt{4 + \frac{\gamma}{2\pi} + \frac{7}{27}\beta^2 \pm \beta}.
\end{equation}
Notice however that the optimal scales $\lambda_\pm$ differ for the plus and minus modes $\phi^\pm$.
The negative mode (counter-rotating vortex) gives the lowest energy for $\beta > 0$ and is shown as the dashed purple curve in the plots.

\subsection{Powers of the versiera}
\label{app:versiera}

Another radially symmetric state is the
``versiera'' state \eqref{eq:versiera-state} centered at the origin,
\begin{equation}
    \phi(z) = \frac{1}{\sqrt{\pi}} \frac{1}{r^2+1},
\end{equation}
which is an NLL state at $\beta=2$.
It however has a divergent potential energy.
More generally, we can take this state to a positive power $a \geq 1$ (and normalize), yielding the state
\begin{equation}
    \phi^a(z) := \sqrt{\frac{2a-1}{\pi}} \frac{1}{(r^2+1)^a}.
\end{equation}
The corresponding density is
\[
    \varrho^a(r) = \frac{2a-1}{\pi} \frac{1}{(r^2+1)^{2a}},
\]
which satisfies $\int_{\R^2} \varrho^a = 1$.
The corresponding magnetic potential is $\bA_{\varrho^a} = A(r)\be_\theta$, with
\[
    A(r) = \frac{1}{r}\left(1-\frac{1}{(r^2+1)^{2a-1}}\right).
\]
Since $\phi^a$ is real, $\bJ_{\phi^a} = 0$.

The kinetic energy is
\[
    \int_{\R^2} \lvert \nabla \phi^a \rvert^2 = \int_{\R^2} \left\lvert \frac{\partial \phi^a}{\partial r} \right\rvert^2 = \frac{2a(2a-1)}{2a+1}.
\]
The potential energy in a harmonic trap diverges logarithmically for $a = 1$,
but for $a > 1$ we get a finite energy contribution
\[
    \int_{\R^2} \Vh \varrho^a = (2a-1) \int_0^\infty \frac{r^2}{(1+r^2)^{2a}} \cdot 2 r \, \mathrm{d}r = \frac{1}{2(a-1)}.
\]
The scalar self-energy is
\begin{align*}
    \int_{\R^2} (\varrho^a)^2 
    &= \frac{1}{\pi}\frac{(2a-1)^2}{4a-1}.
\end{align*}
The magnetic self-energy can be expressed using the digamma function
\begin{equation}
    \psi(z) := \frac{d}{d z}\log \Gamma(z) = \frac{\Gamma'(z)}{\Gamma(z)}.
\end{equation}
Namely, we get
\[
    \int_{\R^2} \lvert \bA_{\varrho^a} \rvert^2 \varrho^a = (2a-1)\vartheta(a),
\]
where we introduce
\begin{equation}
    \vartheta(a) := 2\psi(4a-1) - \psi(2a) - \psi(6a-2).
\end{equation}
This can be shown by evaluating the integral
\begin{align*}
    \int_{\R^2} \lvert \bA_{\varrho^a} \rvert^2 \varrho^a = 2(2a-1) \int_0^\infty \frac{(1-(1+r^2)^{-(2a-1)})^2}{r(1+r^2)^{2a}} \, \mathrm{d}r,
\end{align*}
using a change of variables $t = 1+r^2$, and the digamma function identity
\begin{equation}
    \int_1^\infty \frac{t^{-b} - t^{-a}}{t-1} \, \mathrm{d}t = \psi(b) - \psi(a),
\end{equation}
valid for $a,b\in\C$ with positive real part.

We now consider the rescaled state $\phi^a_\lambda$. At the optimal scale $\lambda = \lambda_*$ (see \eqref{eq:optimal-energy}), we have the total energy 
\begin{align}
    &\cE_{\beta,\gamma,\Vh}\left[\phi^a_{\lambda_*}\right] \nonumber\\
    &= \sqrt{\frac{2(2a-1)}{a-1}}\sqrt{\frac{2a}{2a+1} + \frac{1}{\pi}\frac{2a-1}{4a-1}\gamma + \vartheta(a)\beta^2}.
\end{align}
For fixed $(\beta,\gamma)$, this expression can be numerically optimized over $a > 1$, yielding an optimal state $\phi^{a_*}$. This has been done using Mathematica, and the resulting curve $E = \cE[\phi^{a_*}]$ for $\gamma=2\pi\beta\kappa$ is the thick yellow 
line in the plots.

We note that the Gaussian can be re-obtained by e.g.~setting $a = 2/b$, $\lambda = \sqrt{b/2}$, and taking the limit $b \to 0$. Hence the state $\phi^a$ can (for optimal $a>1$) be expected to have a lower energy than the Gaussian state.

We also note that along the critical line $\gamma = -2\pi\beta$, the optimal energy $\cE_{\beta,\gamma,\Vh}[\phi^a_{\lambda_*}]$ tends to 0 as $\beta$ tends to the critical value $\beta = 2$.
This may indeed be derived analytically using the observations
\begin{align*}
    \psi(z+1) = \psi(z) + \frac{1}{z},
    \qquad
    \psi'(z+1) = \psi'(z) - \frac{1}{z^2},
\end{align*}
which follow from the identity $\Gamma(z+1) = z \Gamma(z)$.

\subsection{Vortex rings}
\label{app:vortexrings}

Finally, we consider a symmetric NLL ``vortex ring'' state of degree (vorticity) $n-1>0$, centered at $w=0$:
\begin{equation}
    \phi(z) = \sqrt{\frac{n}{\pi}} \frac{\bar{z}^{n-1}}{\lvert z \rvert^{2n} + 1}.
\end{equation}
In polar coordinates, $z = re^{i\theta}$, we have
\[
    \phi(z) = \sqrt{\frac{n}{\pi}} \frac{r^{n-1}e^{-i(n-1)\theta}}{r^{2n}+1}.
\]
The corresponding density
\[
    \varrho(r) = \frac{n}{\pi} \frac{r^{2(n-1)}}{(r^{2n}+1)^2}
\]
is radially symmetric and normalized: $\int_{\R^2} \varrho = 1$.
The magnetic-kinetic self-energy has already been computed in \cite[Prop.~3.35 (proof)]{AtaLunNgu-24} using the supersymmetric factorization method:
\begin{align*}
    \cE_{\beta,\gamma=-2\pi\beta,0}[\phi] 
        &= (\beta-2n)^2 \frac{1}{6} \Gamma\left(1+\frac{1}{n}\right) \Gamma\left(3-\frac{1}{n}\right),
\end{align*}
which, with the scalar self-energy
\begin{align*}
    \int_{\R^2} \varrho^2
        &= \frac{n}{6\pi} \Gamma\left(2-\frac{1}{n}\right) \Gamma\left(2+\frac{1}{n}\right),
\end{align*}
yields the total magnetic-kinetic self-energy:
\begin{align*}
    K_1 = \cE_{\beta,\gamma,0}[\phi]
        &= \cE_{\beta,-2\pi\beta,0}[\phi]
        + (2\pi\beta+\gamma) \int_{\R^2} \varrho^2.
\end{align*}
The potential energy is
\[
    U_1 = \int_{\R^2} V_h\varrho 
    = \Gamma\left(1+\frac{1}{n}\right)\Gamma\left(1-\frac{1}{n}\right)
    = \frac{\pi/n}{\sin(\pi/n)},
\]
using the beta function identity
\[
     B(z_1,z_2) = \frac{\Gamma(z_1)\Gamma(z_2)}{\Gamma(z_1+z_2)} = \int_0^\infty \frac{t^{z_1-1}}{(t+1)^{z_1+z_2}} \, \mathrm{d}t,
\]
and Euler's reflection formula,
\[
    \Gamma(1+z)\Gamma(1-z) = \frac{\pi z}{\sin(\pi z)}.
\]
Hence, at the optimal scaling parameter $\lambda=\lambda_*$ (see \eqref{eq:optimal-energy}), the minimal total energy becomes
\begin{align*}
    &\cE_{\beta,\gamma,\Vh}[\phi_{\lambda_*}] \\
    &= \sqrt{\frac{2}{3}} \sqrt{(\beta-2n)^2\left(2-\frac{1}{n}\right) + (2\pi\beta + \gamma)\frac{n+1}{\pi}} \\
    &\qquad \times \sqrt{1-\frac{1}{n}} 
    \,\frac{\pi/n}{\sin(\pi/n)}.
\end{align*}
For $n \in \{2,3,4,5\}$ (vorticity $1,2,3,4$) and $\gamma=2\pi\beta\kappa$, these are the solid purple, orange, green, respectively blue curves in the plots.
Note that indeed $\cE \to 0$ linearly as $\beta \to 2n$ and $\kappa = -1$.


\section{Numerical data using Matlab (unconstrained)}
\label{app:numerics-matlab}

Here we tabulate approximate energies of \eqref{eq:CSS-func} computed in Matlab and optimized using a steepest descent algorithm adopted from \cite{CorDubLunRou-19} to the general CSS equation \eqref{eq:CSS-EL}.
As initial states we have started from the states generated in the work \cite{CorDubLunRou-19} with $\gamma=0$, and adjusted the parameters $(\beta,\gamma)$ to nearby values in order to generate approximate ground states with different vorticities (the number of vortices listed is approximate in non-symmetric cases). In the stable regime away from criticality, convergence to the desired precision is typically achieved after 50-200 iterations of the algorithm. Due to geometrically and topologically induced metastability, the energy converges into a local minimum, and the global minimum may only be found by comparing a set of reasonable vorticities. Some chosen initial states exhibit sufficient metastability that they can produce estimates for excited states well above the ground levels, by stopping the algorithm after a limited number of iterations during which a stable phase is shown. 
Thus, all numerically computed energies are only valid as approximate upper bounds to the true ground-state energy, but may also yield indications of higher resonances. 
At criticality $\kappa=-1$ we have only been able to obtain sufficiently stable and reliable numerics for radially symmetric states.
We have also noted some observations, such as approximate linearity, comparison to analytical 
energies\footnote{Note that the constant $c = 2\sqrt{\pi}/3$ is a numerical fit approximation to that denoted by $e(1,1)/(2\pi)$ in the earlier work \cite{CorDubLunRou-19}.}, 
or cases where we suspect that the state is far from converged.
The resulting energies are shown in the plots using solid markers (colors and shapes indicate numbers of visible vortices; see legend in Fig.~\ref{fig:numerics-rep} (bottom)).


\begin{table}[H]
    \centering
    \begin{tabular}{|l|l|l|l|l|l|l|}
    \hline
        $\beta$  & $\gamma / \pi$ & \#vortices & $\cE[\phi]$ & Remark \\ \hline 
        \hline
        0 & -1.8 & 0 & 0.789 & ~ \\ \hline
        ~ & -1 & 0 & 1.396 & ~ \\ \hline
        ~ & 0 & 0 & 2.000 & exact: 2 \\ \hline
        ~ & 0 & 1 & 4.000 & exact: 4 \\ \hline
        ~ & 2 & 0 & 2.810 & $\ETF(\gamma) \approx 1.886$ \\ \hline
        ~ & 3.18 & 0 & 3.184 & BC: 3.1846 \\ \hline
        ~ & 15.92 & 0 & 5.793 & BC: 5.7920 \\ \hline
        ~ & 20 & 0 & 6.399 & $\ETF(\gamma) \approx 5.963$ \\ \hline
        ~ & 31.84 & 0 & 7.893 & BC: 7.8918 \\ \hline
        ~ & 79.58 & 0 & 12.158 & BC: 12.1578 \\ \hline
        ~ & 159.16 & 0 & 17.024 & BC: 17.0236 \\ \hline
        ~ & 200 & 0 & 19.042 & $\ETF(\gamma) \approx 18.856$ \\ \hline
        ~ & 2000 & 0 & 59.703 & $\ETF(\gamma) \approx 59.628$ \\ \hline
        \hline
        0.1 & -0.2 & 0 & 1.900 & linear \\ \hline
        ~ & -0.2 & 1 & 3.899 & linear? \\ \hline
        ~ & 0 & 0 & 2.003 & ~ \\ \hline
        ~ & 0 & 1 & 3.950 & ~ \\ \hline
        ~ & 0.2 & 0 & 2.100 & linear \\ \hline
        ~ & 0.2 & 1 & 4.001 & ~ \\ \hline
        \hline
        0.5 & -1 & 0 & 1.500 & linear \\ \hline
        ~ & -1 & 1 & 3.500 & linear \\ \hline
        ~ & 0 & 0 & 2.071 & ~ \\ \hline
        ~ & 0 & 1 & 3.774 & ~ \\ \hline
        ~ & 1 & 0 & 2.500 & linear \\ \hline
        ~ & 1 & 1 & 4.030 & ~ \\ \hline
    \end{tabular}
    \caption{Energies computed in Matlab. 
    BC is comparison to Bao--Cai \cite[Table~3.5]{BaoCai-12}
    (note factor 2 and possible rounding errors).} 
    \label{tab:matlab-energies-0}
\end{table}

\begin{table}[H]
    \centering
    \begin{tabular}{|l|l|l|l|l|l|l|}
    \hline
        $\beta$  & $\gamma / \pi$ & \#vortices & $\cE[\phi]$ & Remark \\ \hline 
        \hline
        1 & -2 & 0 & 1.003 & linear? \\ \hline
        ~ & -2 & 1 & 3.013 & linear? \\ \hline
        ~ & -1.8 & 0 & 1.204 & ~ \\ \hline
        ~ & -1.8 & 1 & 3.078 & ~ \\ \hline
        ~ & 0 & 0 & 2.268 & $\ETF(2\pi c) \approx 2.050$ \\ \hline
        ~ & 0 & 1 & 3.607 & ~ \\ \hline
        ~ & 1 & 0 & 2.662 & ~ \\ \hline
        ~ & 2 & 0 & 3.000 & linear \\ \hline
        ~ & 2 & 1 & 4.120 & ~ \\ \hline
        \hline
        1.5 & -3 & 0 & 0.631 & unconverged? \\ \hline
        ~ & -3 & 1 & 2.542 & unconverged? \\ \hline
        ~ & 0 & 0 & 2.560 & ~ \\ \hline
        ~ & 0 & 1 & 3.505 & ~ \\ \hline
        ~ & 3 & 0 & 3.500 & linear \\ \hline
        ~ & 3 & 1 & 4.265 & ~ \\ \hline
        \hline
        1.9 & -3.8 & 0 & 0.188 & unconverged? \\ \hline
    \end{tabular}
    \caption{Energies computed in Matlab.} 
    \label{tab:matlab-energies-1}
\end{table}

\begin{table}[H]
    \centering
    \begin{tabular}{|l|l|l|l|l|l|l|}
    \hline
        $\beta$  & $\gamma / \pi$ & \#vortices & $\cE[\phi]$ & Remark \\ \hline 
        \hline
        2 & -4 & 1 & 2.024 & linear? \\ \hline
        ~ & -3.6 & 0 & 1.081 & ~ \\ \hline
        ~ & -3.6 & 1 & 2.227 & ~ \\ \hline
        ~ & -2 & 0 & 2.145 & ~ \\ \hline
        ~ & -2 & 1 & 2.841 & ~ \\ \hline
        ~ & 0 & 0 & 2.916 & $\ETF(4\pi c) \approx 2.899$ \\ \hline
        ~ & 0 & 1 & 3.474 & ~ \\ \hline
        ~ & 2 & 0 & 3.504 & ~ \\ \hline
        ~ & 2 & 1 & 4.001 & ~ \\ \hline
        ~ & 4 & 0 & 4.000 & linear \\ \hline
        ~ & 4 & 1 & 4.462 & ~ \\ \hline
        \hline
        2.5 & -5 & 0 & 0.736 & unconverged? \\ \hline
        ~ & -5 & 1 & 1.553 & ~ \\ \hline
        ~ & 0 & 0 & 3.313 & ~ \\ \hline
        ~ & 0 & 1 & 3.517 & ~ \\ \hline
        ~ & 5 & 0 & 4.500 & linear \\ \hline
        ~ & 5 & 1 & 4.703 & ~ \\ \hline
    \end{tabular}
    \caption{Energies computed in Matlab.} 
    \label{tab:matlab-energies-2}
\end{table}

\begin{table}[H]
    \centering
    \begin{tabular}{|l|l|l|l|l|l|l|}
    \hline
        $\beta$  & $\gamma / \pi$ & \#vortices & $\cE[\phi]$ & Remark \\ \hline 
        \hline
        3 & -6 & 0 & 1.393 & ~ \\ \hline
        ~ & -6 & 1 & 1.000 & linear \\ \hline
        ~ & -5.4 & 0 & 1.860 & ~ \\ \hline
        ~ & -5.4 & 1 & 1.506 & ~ \\ \hline
        ~ & -5 & 1 & 1.761 & ~ \\ \hline
        ~ & 0 & 0 & 3.736 & ~ \\ \hline
        ~ & 0 & 1 & 3.631 & $\ETF(6\pi c) \approx 3.550$ \\ \hline
        ~ & 6 & 0 & 5.000 & linear \\ \hline
        ~ & 6 & 1 & 4.983 & ~ \\ \hline
        \hline
        3.5 & -7 & 1 & 0.504 & linear? \\ \hline
    \end{tabular}
    \caption{Energies computed in Matlab.} 
    \label{tab:matlab-energies-3}
\end{table}

\begin{table}[H]
    \centering
    \begin{tabular}{|l|l|l|l|l|l|l|}
    \hline
        $\beta$  & $\gamma / \pi$ & \#vortices & $\cE[\phi]$ & Remark \\ \hline 
        \hline
        4 & -8 & 0 & 2.528 & ~ \\ \hline
        ~ & -8 & 1 & 0.090 & unconverged? \\ \hline
        ~ & -7.2 & 0 & 2.829 & ~ \\ \hline
        ~ & -7.2 & 1 & 1.332 & ~ \\ \hline
        ~ & -7.2 & 2 & 1.708 & ~ \\ \hline
        ~ & -6 & 1 & 2.077 & ~ \\ \hline
        ~ & -6 & 2 & 2.450 & ~ \\ \hline
        ~ & -4 & 1 & 2.901 & ~ \\ \hline
        ~ & -4 & 2 & 3.240 & ~ \\ \hline
        ~ & -2 & 1 & 3.525 & ~ \\ \hline
        ~ & -2 & 2 & 3.835 & ~ \\ \hline
        ~ & 0 & 0 & 4.631 & ~ \\ \hline
        ~ & 0 & 1 & 4.046 & $\ETF(8\pi c) \approx 4.099$ \\ \hline
        ~ & 0 & 2 & 4.335 & ~ \\ \hline
        ~ & 0 & 3 & 4.536 & ~ \\ \hline
        ~ & 8 & 0 & 6.000 & linear \\ \hline
        ~ & 8 & 1 & 5.634 & ~ \\ \hline
        ~ & 8 & 2 & 5.878 & ~ \\ \hline
        ~ & 8 & 3 & 6.102 & ~ \\ \hline
        ~ & 16 & 1 & 6.839 & ~ \\ \hline
        ~ & 16 & 2 & 7.062 & ~ \\ \hline
        ~ & 2000 & 0 & 59.840 & ~ \\ \hline
        ~ & 2000 & 1 & 59.890 & ~ \\ \hline
        ~ & 2000 & 2 & 59.970 & ~ \\ \hline
        \hline
        4.5 & -9 & 1 & 0.535 & linear? \\ \hline
    \end{tabular}
    \caption{Energies computed in Matlab.} 
    \label{tab:matlab-energies-4}
\end{table}

\begin{table}[H]
    \centering
    \begin{tabular}{|l|l|l|l|l|l|l|}
    \hline
        $\beta$  & $\gamma / \pi$ & \#vortices & $\cE[\phi]$ & Remark \\ \hline 
        \hline
        5 & -10 & 0 & 3.546 & ~ \\ \hline
        ~ & -10 & 1 & 1.076 & linear? \\ \hline
        ~ & -9 & 1 & 1.850 & ~ \\ \hline
        ~ & -9 & 2 & 1.679 & ~ \\ \hline
        ~ & -9 & 3 & 1.805 & ~ \\ \hline
        ~ & 0 & 1 & 4.649 & ~ \\ \hline
        ~ & 0 & 2 & 4.641 & $\ETF(10\pi c) \approx 4.583$ \\ \hline
        ~ & 0 & 3 & 4.826 & ~ \\ \hline
        ~ & 0 & 4 & 5.100 & ~ \\ \hline
        ~ & 10 & 1 & 6.377 & ~ \\ \hline
        ~ & 10 & 2 & 6.402 & ~ \\ \hline
        ~ & 10 & 3 & 6.572 & ~ \\ \hline
        ~ & 10 & 4 & 6.817 & ~ \\ \hline
    \end{tabular}
    \caption{Energies computed in Matlab.} 
    \label{tab:matlab-energies-5}
\end{table}

\begin{table}[H]
    \centering
    \begin{tabular}{|l|l|l|l|l|l|l|}
    \hline
        $\beta$  & $\gamma / \pi$ & \#vortices & $\cE[\phi]$ & Remark \\ \hline 
        \hline
        6 & -11.8 & 2 & 0.715 & approx. ring \\ \hline
        ~ & -10.8 & 0 & 4.761 & ~ \\ \hline
        ~ & -10.8 & 1 & 2.690 & ~ \\ \hline
        ~ & -10.8 & 2 & 1.708 & approx. ring \\ \hline
        ~ & -10.8 & 3 & 1.790 & ~ \\ \hline
        ~ & -10.8 & 4 & 2.175 & ~ \\ \hline
        ~ & -10 & 2 & 2.183 & ~ \\ \hline
        ~ & -10 & 3 & 2.246 & ~ \\ \hline
        ~ & -6 & 1 & 4.145 & ~ \\ \hline
        ~ & -6 & 2 & 3.671 & ~ \\ \hline
        ~ & -6 & 3 & 3.688 & ~ \\ \hline
        ~ & -6 & 4 & 3.959 & ~ \\ \hline
        ~ & 0 & 0 & 6.510 & ~ \\ \hline
        ~ & 0 & 1 & 5.379 & ~ \\ \hline
        ~ & 0 & 2 & 5.066 & $\ETF(12\pi c) \approx 5.021$ \\ \hline
        ~ & 0 & 3 & 5.078 & ~ \\ \hline
        ~ & 0 & 4 & 5.314 & ~ \\ \hline
        ~ & 12 & 0 & 8.000 & linear \\ \hline
        ~ & 12 & 1 & 7.183 & ~ \\ \hline
        ~ & 12 & 2 & 6.992 & ~ \\ \hline
        ~ & 12 & 3 & 7.021 & ~ \\ \hline
        ~ & 12 & 4 & 7.220 & ~ \\ \hline
        ~ & 2000 & 0 & 60.011 & ~ \\ \hline
        ~ & 2000 & 1 & 59.985 & ~ \\ \hline
        ~ & 2000 & 2 & 60.025 & ~ \\ \hline
        ~ & 2000 & 3 & 60.081 & ~ \\ \hline
        ~ & 2000 & 4 & 60.145 & ~ \\ \hline
    \end{tabular}
    \caption{Energies computed in Matlab.} 
    \label{tab:matlab-energies-6}
\end{table}

\begin{table}[H]
    \centering
    \begin{tabular}{|l|l|l|l|l|l|l|}
    \hline
        $\beta$  & $\gamma / \pi$ & \#vortices & $\cE[\phi]$ & Remark \\ \hline 
        \hline
        7 & -13 & 3 & 1.561 & ~ \\ \hline
        ~ & -12.6 & 1 & 3.625 & ~ \\ \hline
        ~ & -12.6 & 2 & 2.127 & 2 or 4 vortices (ring) \\ \hline
        ~ & -12.6 & 3 & 1.824 & ~ \\ \hline
        ~ & -12.6 & 4 & 1.915 & ~ \\ \hline
        ~ & 0 & 1 & 6.188 & ~ \\ \hline
        ~ & 0 & 2 & 5.619 & ~ \\ \hline
        ~ & 0 & 3 & 5.415 & $\ETF(14\pi c) \approx 5.423$ \\ \hline
        ~ & 0 & 4 & 5.518 & ~ \\ \hline
        ~ & 14 & 1 & 8.033 & ~ \\ \hline
        ~ & 14 & 2 & 7.652 & ~ \\ \hline
        ~ & 14 & 3 & 7.522 & ~ \\ \hline
        ~ & 14 & 4 & 7.614 & ~ \\ \hline
    \end{tabular}
    \caption{Energies computed in Matlab.} 
    \label{tab:matlab-energies-7}
\end{table}

\begin{table}[H]
    \centering
    \begin{tabular}{|l|l|l|l|l|l|l|}
    \hline
        $\beta$  & $\gamma / \pi$ & \#vortices & $\cE[\phi]$ & Remark \\ \hline 
        \hline
        8 & -15 & 3 & 1.826 & 3 or 6 vortices (ring) \\ \hline
        ~ & -15 & 4 & 1.566 & ~ \\ \hline
        ~ & -15 & 6  & 1.596 & ~ \\ \hline
        ~ & -14.4 & 0 & 6.721 & ~ \\ \hline
        ~ & -14.4 & 1 & 4.590 & ~ \\ \hline
        ~ & -14.4 & 3 & 2.145 & 3 or 6 vortices (ring) \\ \hline
        ~ & -14.4 & 4 & 1.951 & $\ETF(1.6\pi c) \approx 1.833$ \\ \hline
        ~ & -14.4 & 6 & 1.993 & ~ \\ \hline
        ~ & 0 & 0 & 8.440 & ~ \\ \hline
        ~ & 0 & 1 & 7.049 & ~ \\ \hline
        ~ & 0 & 2 & 6.274 & ~ \\ \hline
        ~ & 0 & 3 & 5.861 & ~ \\ \hline
        ~ & 0 & 4 & 5.792 & $\ETF(16\pi c) \approx 5.797$ \\ \hline
        ~ & 16 & 0 & 10.000 & linear \\ \hline
        ~ & 16 & 1 & 8.915 & ~ \\ \hline
        ~ & 16 & 2 & 8.370 & ~ \\ \hline
        ~ & 16 & 3 & 8.086 & ~ \\ \hline
        ~ & 16 & 4 & 8.051 & $\ETF(32\pi c) \approx 8.199$ \\ \hline
        ~ & 16 & 5 & 8.190 & not symmetric \\ \hline
    \end{tabular}
    \caption{Energies computed in Matlab.} 
    \label{tab:matlab-energies-8}
\end{table}

\begin{table}[H]
    \centering
    \begin{tabular}{|l|l|l|l|l|l|l|}
    \hline
        $\beta$  & $\gamma / \pi$ & \#vortices & $\cE[\phi]$ & Remark \\ \hline 
        \hline
        9 & -16.2 & 1 & 5.582 & ~ \\ \hline
        ~ & -16.2 & 2 & 3.747 & 2 or 4 vortices \\ \hline
        ~ & -16.2 & 6 & 2.216 & 6 or 9 vortices \ \\ \hline
        ~ & -16.2 & 8 & 2.056 & 4 or 8 vortices \\ \hline
        ~ & 0 & 1 & 7.944 & ~ \\ \hline
        ~ & 0 & 2 & 7.011 & 2 or 4 vortices \\ \hline
        ~ & 0 & 4 & 6.160 & $\ETF(18\pi c) \approx 6.149$ \\ \hline
        ~ & 0 & 6 & 6.298 & ~ \\ \hline
        ~ & 18 & 1 & 9.819 & ~ \\ \hline
        ~ & 18 & 2 & 9.136 & ~ \\ \hline
        ~ & 18 & 4 & 8.544 & ~ \\ \hline
        ~ & 18 & 6 & 8.690 & ~ \\ \hline
    \end{tabular}
    \caption{Energies computed in Matlab.} 
    \label{tab:matlab-energies-9}
\end{table}

\begin{table}[H]
    \centering
    \begin{tabular}{|l|l|l|l|l|l|l|}
    \hline
        $\beta$  & $\gamma / \pi$ & \#vortices & $\cE[\phi]$ & Remark \\ \hline 
        \hline
        10 & -18 & 0 & 8.602 & fast collapsing ring \\ \hline
        ~ & -18 & 1 & 6.588 & ~ \\ \hline
        ~ & -18 & 3 & 3.171 & 3 or 6 vortices (ring) \\ \hline
        ~ & -18 & 6 & 2.175 & $\ETF(2\pi c) \approx 2.050$ \\ \hline
        ~ & -18 & 8 & 2.263 & ~ \\ \hline
        ~ & -18 & 9 & 2.296 & ~ \\ \hline
        ~ & -10 & 6 & 4.702 & ~ \\ \hline
        ~ & -10 & 8 & 4.804 & ~ \\ \hline
        ~ & 0 & 0 & 10.392 & ~ \\ \hline
        ~ & 0 & 1 & 8.862 & ~ \\ \hline
        ~ & 0 & 2 & 7.808 & ~ \\ \hline
        ~ & 0 & 3 & 7.040 & ~ \\ \hline
        ~ & 0 & 6 & 6.543 & $\ETF(20\pi c) \approx 6.482$ \\ \hline
        ~ & 0 & 8 & 6.619 & ~ \\ \hline
        ~ & 10 & 4 & 7.967 & ~ \\ \hline
        ~ & 20 & 1 & 10.741 & ~ \\ \hline
        ~ & 20 & 3 & 9.387 & ~ \\ \hline
        ~ & 20 & 4 & 9.095 & ~ \\ \hline
        ~ & 20 & 6 & 9.066 & $\ETF(40\pi c) \approx 9.167$ \\ \hline
        ~ & 40 & 4 & 10.979 & ~ \\ \hline
        ~ & 80 & 4 & 13.959 & ~ \\ \hline
        ~ & 160 & 4 & 18.473 & ~ \\ \hline
        ~ & 200 & 4 & 20.344 & $\ETF(220\pi c) \approx 21.498$ \\ \hline
        ~ & 2000 & 4 & 60.277 & $\ETF(2020\pi c) \approx 65.141$ \\ \hline
    \end{tabular}
    \caption{Energies computed in Matlab.} 
    \label{tab:matlab-energies-10}
\end{table}

\begin{table}[H]
    \centering
    \begin{tabular}{|l|l|l|l|l|l|l|}
    \hline
        $\beta$  & $\gamma / \pi$ & \#vortices & $\cE[\phi]$ & Remark \\ \hline 
        \hline
        50 & -90 & 40 & 4.721 & $\ETF(10\pi c) \approx 4.583$ \\ \hline
        ~ & -50 & 40 & 10.370 & $\ETF(50\pi c) \approx 10.249$ \\ \hline
        ~ & 0 & 40 & 14.502 & $\ETF(100\pi c) \approx 14.494$ \\ \hline
        ~ & 100 & 40 & 20.228 & $\ETF(200\pi c) \approx 20.497$ \\ \hline
        ~ & 200 & 40 & 24.552 & $\ETF(300\pi c) \approx 25.104$ \\ \hline
        \hline
        100 & -180 & 88 & 6.654 & $\ETF(20\pi c) \approx 6.482$ \\ \hline
        ~ & -100 & 88 & 14.643 & $\ETF(100\pi c) \approx 14.494$ \\ \hline
        ~ & 0 & 88 & 20.499 & $\ETF(200\pi c) \approx 20.497$ \\ \hline
        ~ & 200 & 87 & 28.612 & $\ETF(400\pi c) \approx 28.987$ \\ \hline
        ~ & 400 & 88 & 34.748 & $\ETF(600\pi c) \approx 35.502$ \\ \hline
    \end{tabular}
    \caption{Energies computed in Matlab.} 
    \label{tab:matlab-energies-50}
\end{table}

\section{Numerical data using Mathematica (NLL constraint)}
\label{app:numerics-mathematica}

Finally, we use the above analytical solutions to the Wronskian problem for $\beta = 2n \in \{4,6,8,10\}$ to estimate the lowest energy of corresponding NLL states with relaxed radial constraints on the density (a macroscopic, averaged symmetry is still expected due to the radial potential).
For $d := \deg W \in \{2,3,4\}$ and $n=5$, $d=5$, 
we place the vortices in a circle of radius $R \ge 0$:
$w_k = R e^{i2\pi k/d}$, 
while for $n=4, d=5$ and $n=5, d=6$ we put one at the origin and the remaining in a circle; 
compare \eqref{eq:wronsk-beta8-deg3}-\eqref{eq:wronsk-beta10-deg6}.
Further, we choose $\eta=0$ 
due to symmetry
(expecting symmetric configurations to lower the energy in the harmonic potential),
and then numerically optimize the energy
\eqref{eq:NLL-energy}
over the remaining parameters $\lambda > 0$ and $R \ge 0$.
This optimization is done in Mathematica by first scanning for reasonable initial values in a rough energy plot and then iteratively rescaling the state as in \eqref{eq:total-energy}--\eqref{eq:optimal-scaling} in order that the kinetic and potential energies eventually match (convergence towards a fixed point).
At given parameter values, the energy is computed by numerical integration of the analytical density of
\eqref{eq:uPQ-state} in \eqref{eq:NLL-energy} on a finite domain containing almost all of the mass.
The resulting minimal energies are shown in the plots using hollow markers (colors and shapes match the previous vortex numbers; see legend in Fig.~\ref{fig:numerics-rep} (bottom)).
For cases with $n=5$, $d=7,8$ (yielding higher energies) we refer to \cite{Getzner-26}.

\begin{table}[H]
    \centering
    \begin{tabular}{|l|l|l|l|l|l|l|}
    \hline
        $\beta$  & $\deg W$  & $\gamma / \pi$ & $R$ & $\lambda^{-1}$ & $\cE[\phi_{P,Q}]$ & Remark \\ \hline 
        \hline
         4 & 1 & -7.2 & 0 & 0.8944 & 1.4049 & ~ \\ \hline
         ~ & 1 & -6 & 0 & 1.4142 & 2.2214 & ~ \\ \hline
         ~ & 1 & -4 & 0 & 2.0000 & 3.1416 & ~ \\ \hline
        ~ & 1 & 0 & 0 & 2.8284 & 4.4428 & ~ \\ \hline
        ~ & 1 & 8 & 0 & 4.0000 & 6.2832 & ~ \\ \hline 
        ~ & 2 & -7.2 & 0.6054 & 0.7265 & 2.7760 & ~ \\ \hline
        ~ & 2 & -6 & 0.6860 & 1.0290 & 4.4693 & ~ \\ \hline
        ~ & 2 & -4 & 0.8774 & 1.1965 & 5.8111 & ~ \\ \hline
        ~ & 2 & 0 & 1.1519 & 1.3823 & 7.6549 & ~ \\ \hline 
        ~ & 2 & 8 & 1.5034 & 1.6108 & 10.3610 & ~ \\ \hline 
        \hline
        6 & 2 & -10.8 & 0 & 1.8724 & 1.7661 & ~ \\ \hline
        ~ & 2 & -6 & 0 & 6.2614 & 3.9487 & ~ \\ \hline
        ~ & 2 & 0 & 0 & 10.5320 & 5.5838 & ~ \\ \hline
        ~ & 2 & 12 & 0 & 17.7194 & 7.8948 & ~ \\ \hline
        ~ & 3 & -10.8 & 0.5055 & 2.0858 & 2.3043 & ~ \\ \hline
        ~ & 3 & -6 & 0.8344 & 3.9167 & 4.5982 & ~ \\ \hline
        ~ & 3 & 0 & 0.9697 & 5.6420 & 6.4626 & ~ \\ \hline
        ~ & 3 & 12 & 1.1428 & 8.0949 & 9.1033 & ~ \\ \hline
        ~ & 4 & -10.8 & 0.8861 & 0.9747 & 3.7769 & errors \\ \hline
        ~ & 4 & -6 & 1.3621 & 1.5437 & 7.8009 & ~ \\ \hline
        ~ & 4 & 0 & 1.5355 & 2.1498 & 10.8994 & ~ \\ \hline
        ~ & 4 & 12 & 1.9367 & 2.4209 & 14.3487 & ~ \\ \hline
    \end{tabular}
    \caption{Energies for optimized NLL states.} 
    \label{tab:NLL-energies-4}
\end{table}

\begin{table}[H]
    \centering
    \begin{tabular}{|l|l|l|l|l|l|l|}
    \hline
        $\beta$  & $\deg W$  & $\gamma / \pi$ & $R$ & $\lambda^{-1}$ & $\cE[\phi_{P,Q}]$ & Remark \\ \hline 
        \hline
        8 & 3 & -14.4 & 0 & 4.0000 & 2.2214 & errors \\ \hline
        ~ & 3 & -14 & 0 & 5.0000 & 2.4836 & ~ \\ \hline
        ~ & 3 & -8 & 0 & 20.0010 & 4.9673 & ~ \\ \hline
        ~ & 3 & 0 & 0 & 40.0021 & 7.0247 & ~ \\ \hline
        ~ & 3 & 8 & 0 & 60.0031 & 8.6034 & ~ \\ \hline
        ~ & 3 & 16 & 0 & 80.0042 & 9.9334 & ~ \\ \hline
        ~ & 4 & -14.4 & 0.5859 & 3.4455 & 2.1411 & (a) \\ \hline 
        ~ & 4 & -14 & 0.6235 & 4.0390 & 2.3895 & (b) \\ \hline 
        ~ & 4 & -8 & 0.8845 & 11.3902 & 4.7762 & (c) \\ \hline 
        ~ & 4 & 0 & 1.0640 & 19.0020 & 6.7574 & (d) \\ \hline 
        ~ & 4 & 8 & 1.1502 & 26.4183 & 8.2982 & ~ \\ \hline
        ~ & 4 & 16 & 1.2622 & 32.0333 & 9.5480 & ~ \\ \hline
        ~ & 5 & -14.4 & 0.9204 & 2.5415 & 2.6768 & ~ \\ \hline
        ~ & 5 & -14 & 0.8534 & 2.9129 & 2.7407 & ~ \\ \hline
        ~ & 5 & -8 & 1.1760 & 5.9547 & 5.4152 & errors \\ \hline
        ~ & 5 & 0 & 1.4431 & 8.3299 & 7.6605 & ~ \\ \hline
        ~ & 5 & 8 & 1.5369 & 10.4986 & 9.2824 & ~ \\ \hline
        ~ & 5 & 16 & 1.6036 & 12.5713 & 10.7563 & ~ \\ \hline
        ~ & 6 & -14.4 & 1.1386 & 1.1238 & 4.1588 & ~ \\ \hline
        ~ & 6 & -8 & 1.7287 & 1.8090 & 8.6607 & ~ \\ \hline
        ~ & 6 & 0 & 2.0726 & 2.2416 & 11.8231 & ~ \\ \hline
        ~ & 6 & 8 & 2.3074 & 2.5523 & 14.1536 & ~ \\ \hline
        ~ & 6 & 16 & 2.4915 & 2.8058 & 16.0680 & ~ \\ \hline
    \end{tabular}
    \caption{Energies for optimized NLL states.} 
    \label{tab:NLL-energies-8}
\end{table}

\begin{table}[H]
    \centering
    \begin{tabular}{|l|l|l|l|l|l|l|}
    \hline
        $\beta$  & $\deg W$  & $\gamma / \pi$ & $R$ & $\lambda^{-1}$ & $\cE[\phi_{P,Q}]$ & Remark \\ \hline 
        \hline
        10 & 4 & -18 & 0 & 9.0009 & 2.7043 & ~ \\ \hline
        ~ & 4 & -10 & 0 & 67.2721 & 6.0469 & ~ \\ \hline
        ~ & 4 & 0 & 0 & 160.0000 & 8.5517 & ~ \\ \hline
        ~ & 4 & 10 & 0 & 265.6042 & 10.4736 & ~ \\ \hline
        ~ & 4 & 20 & 0 & 380.5465 & 12.0939 & ~ \\ \hline
        ~ & 5 & -18 & 0.6651 & 6.7254 & 2.4356 & ~ \\ \hline
        ~ & 5 & -10 & 0.9946 & 33.6281 & 5.4461 & ~ \\ \hline
        ~ & 5 & 0 & 1.1914 & 66.8896 & 7.7077 & ~ \\ \hline
        ~ & 5 & 10 & 1.3143 & 100.5338 & 9.4350 & ~ \\ \hline
        ~ & 5 & 20 & 1.4139 & 134.5770 & 10.8960 & ~ \\ \hline
        ~ & 6 & -18 & 0.7579 & 4.7441 & 2.2424 & ~ \\ \hline
        ~ & 6 & -10 & 1.1334 & 15.8700 & 5.0128 & ~ \\ \hline
        ~ & 6 & 0 & 1.3481 & 26.7051 & 7.0867 & ~ \\ \hline
        ~ & 6 & 10 & 1.4921 & 36.2161 & 8.6764 & ~ \\ \hline
        ~ & 6 & 20 & 1.6036 & 44.9640 & 10.0151 & ~ \\ \hline
    \end{tabular}
    \caption{Energies for optimized NLL states.} 
    \label{tab:NLL-energies-10}
\end{table}

Remarks:

Some of the numerical integrations reported errors regardless of precision and may therefore be less reliable.

(a) 
The energy at the lowest radial vortex ring here ($n=4$) is $\pi/\sqrt{2} \approx 2.2214$.

(b) 
The energy at the lowest radial vortex ring here ($n=4$) is $\sqrt{5/2}\pi/2 \approx 2.4836$.

(c) 
The energy at the lowest radial vortex ring here ($n=4$) is $\sqrt{5/2}\pi \approx 4.9673$.

(d) 
The energy at the lowest radial vortex ring here ($n=3$) is $8\sqrt{53}\pi/27 \approx 6.7766$.

\def\MR#1{} 



\end{document}